\documentclass{aa}
\usepackage{epsfig}
\usepackage{graphics}
\usepackage{subfigure}
\usepackage{natbib}
\usepackage{amsmath}
\usepackage{enumerate}
\usepackage{amssymb}
\usepackage{upgreek}

\begin{document}

\title{Particle acceleration in the superwinds of starburst galaxies}

\author{G. E. Romero\inst{1,2}, A.L. M\"uller\inst{1,3,4}, M. Roth\inst{3}}
  
\institute{Instituto Argentino de Radioastronom\'{\i}a (IAR, CCT La Plata, CONICET/CIC), C.C.5, (1984) Villa Elisa, Buenos Aires, Argentina \and Facultad de Ciencias Astron\'omicas y Geof\'{\i}sicas, Universidad Nacional de La Plata, Paseo del Bosque s/n, 1900, La Plata, Argentina \and Institute for Nuclear Physics (IKP), Karlsruhe Institute of Technology (KIT), Germany \and Instituto de Tecnolog\'{\i}as en Detecci\'on y Astropart\'{\i}culas (CNEA, CONICET, UNSAM), Buenos Aires, Argentina}

\offprints{G.E. Romero \\ \email{romero@iar-conicet.gov.ar}}

\titlerunning{Particle acceleration in superwinds}

\authorrunning{Romero, et al.}

\abstract
{Starbursts are galaxies undergoing massive episodes of star formation. The combined effect of stellar winds from hot stars and supernova explosions creates a high-temperature cavity in the nuclear region of these objects. The very hot gas expands adiabatically and escapes from the galaxy creating a superwind which sweeps matter from the galactic disk. The superwind region in the halo is filled with a multi-phase gas with hot, warm, cool, and relativistic components. }
{The shocks associated with the superwind of starbursts and the turbulent gas region of the bubble inflated by them might accelerate cosmic rays up to high energies. In this work we calculate the cosmic ray production associated with the superwind using parameters that correspond to the nearby southern starburst galaxy NGC 253, which has been suggested as a potential accelerator of ultra-high-energy cosmic rays. }
{We evaluate the efficiency of both diffusive shock acceleration (DSA) and stochastic diffusive acceleration (SDA) in the superwind of NGC 253. We estimate the distribution of both hadrons and leptons and calculate the corresponding spectral energy distributions of photons. The electromagnetic radiation can help to discriminate between the different scenarios analyzed.}
{We find that the strong mass load of the superwind, recently determined through ALMA observations, strongly attenuates the efficiency of DSA in NGC 253, whereas SDA is constrained by the age of the starburst.}
{We conclude that NGC 253 and similar starbursts can only  accelerate iron nuclei beyond  $\sim10^{18}$ eV under very special conditions.  If the central region of the galaxy harbors a starved supermassive black hole of $\sim10^6$ $M_{\odot}$, as suggested by some recent observations, a contribution in the range $10^{18}-10^{19}$ eV can be present for accretion rates $\dot{m}\sim10^{-3}$ in Eddington units. Shock energies of the order of 100 EeV might only be possible if very strong magnetic field amplification occurs close to the superwind.}

 
\maketitle

\section{Introduction}\label{Intro}

Intense star formation in galaxies results in the ejection of gas into the surrounding medium. This outflow, driven by the collective effect of the stellar winds of hot stars and supernova explosions,  plays an important role in galaxy formation and evolution. In particular, it can quench star formation by depleting the gas from the central regions of the galaxy. The outflow thrusted by the starburst constitutes a self-regulatory mechanism that prevents the stellar mass of the galaxy from growing too much. When the gas escapes, star formation ceases, and the outflow stops. Star formation can then be reignited by fresh gas inflow. This might occur as a consequence of gravitational interactions with nearby galaxies or when part of the expelled gas that is still gravitationally bound precipitates back as high-velocity clouds.  On the other hand, the outflow transports metals to the intergalactic space and gas to the halo.  As a consequence, the intergalactic medium is chemically enriched to the point that a significant fraction of all metals ever created in the Universe lie outside galaxies (e.g., \citealt{Pagel2002}).

The existence of galactic superwinds remained speculative for a long time. In the last 20 years a battery of multi-wavelength observations of nearby galaxies such as M82 and NGC 253 has revealed many aspects of the nature, frequency, structure, and role of these winds. All local starburst galaxies appear to have superwinds (e.g., \citealt{veilleux2005}) and since star formation increases towards high redshifts, superwinds seem to be ubiquitous in the Universe. 

The basic mechanism that creates a superwind in a starburst was established by \cite{chevalier1985} long ago, and since then the basic model has been supported by observations (e.g., \citealt{heckman1990}) and complemented with detailed simulations (e.g., \citealt{strickland2000}, \citealt{cooper2008}). The large number of core-collapse supernovae in starbursts results in the merge of supernova remnants before they have had time to lose their energy. Shocks formed in these collisions thermalize the energy released by the explosions creating a cavity filled with hot \mbox{($T\sim 10^8$ K)} gas that is unbound by the gravitational potential because its temperature is greater than the local escape temperature. The hot gas expands adiabatically, becomes supersonic at the edge of the starburst region, and finally escapes the system sweeping up cooler and denser gas from the disk. The outflow is then multi-phased with components of different temperatures. If the velocity of the expanding gas exceeds the escape velocity of the galaxy, part of the swept medium is transferred to the intergalactic space. 

The hot gas expanding into the halo creates an X-ray-emitting region surrounded by swept warm \mbox{($T\sim10^{4}$ K)} gas that radiates H$\alpha$ lines. The superwind activity can be traced up to distances of $\sim10$ kpc from the disk of edge-on galaxies (e.g., \citealt{strickland2002}). Radio observations of nearby starbursts show up an extended halo of synchrotron radiation (e.g., \citealt{heesen2009a}). Such observations support the idea that particle acceleration and transport can be associated with the superwind. 

Cosmic ray acceleration in galactic superwinds has already been discussed by \citet{Jokipii1985} and \citet{bustard2017} for the case of our Galaxy, and by \citet{anchordoqui1999} for the nearby starburst NGC 253.  In addition, it has been suggested that the extended gamma-ray bubbles observed by the \textit{Fermi} satellite \citep{su2010} might be the radiative signature of particles accelerated in the termination shock of the large-scale wind produced by a star-forming episode in the central region of the Galaxy \citep{lacki2014}.  All these models invoke diffusive shock acceleration at the termination shock of the outflows. \citet{anchordoqui1999}, in particular, suggested that nearby starburst galaxies might be responsible for the acceleration of heavy nuclei up to ultra-high energies ($\sim 10^{20}$ eV). They based their proposal on three facts: 1) The large extent of the superwind region of nearby starbursts can accommodate high-energy cosmic rays with equipartition magnetic fields; 2) the photon density in the halo, contrary to the central region of these galaxies, is sufficiently low as to prevent photo-disintegration of the nuclei during the acceleration by diffusive processes; and 3)  the high metallicity of the wind provides a pool of nuclei from which a Fermi type I mechanism can operate (e.g., \citealt{drury1983}, \citealt{protheroe1999}) to yield high-energy cosmic rays of heavy composition. At the time of publication by  \cite{anchordoqui1999}, little was known about the actual superwind of NGC 253 so their discussion was not informed with realistic estimates of the physical conditions in the outflow region of this galaxy. 

In the last 20 years, many observations across the entire electromagnetic spectrum of NGC 253 and other nearby galaxies have allowed for better characterization of starburst-driven superwinds and their interaction with the intergalactic medium. Also, the Pierre Auger Observatory (PAO) has detected a change in the composition of cosmic rays towards heavy nuclei at high energies. In particular, the most recent measurements in combination
with post-Large Hadron Collider (LHC) hadronic models show the absence or a small fraction of both protons and iron at $E > 10$ EeV (\citealt{ Auger2014}, \citealt{Auger2017a}). All this points in the direction of cosmic accelerators with high metallicity.  The Telescope Array (TA) data are consistent with protons for pre-LHC models, but do not have sensitivity to distinguish protons from intermediate nuclei at the same energy.  A joint analysis of both experiments has shown a consistency of the experimental data on composition between TA and PAO within estimated errors \citep{TA20016}. According to \cite{semikoz2016} a possible solution consistent with the  existing data could be that ``cosmic rays at $E > 40$ EeV are largely composed of intermediate-mass nuclei, and their deflections prevent us from finding sources by correlating arrival directions with the source positions''. Also, differences in the cut-off at the highest energies measured by both experiments might be caused by different nearby sources dominating the Northern and Southern skies. In fact, TA has recently claimed to have detected a ``hotspot''  in the Northern hemisphere using the five-year data recorded up to May 4, 2013 \citep{TA2014}. The hotspot was a cluster of 19 events with energies $> 57$ EeV occupying a circle of $\sim 20\deg$ in radius close to the Ursa Major cluster of galaxies. The pre-trial statistical significance of the hotspot is $5.1\sigma$, with the post-trial probability of being by chance in an isotropic cosmic ray sky of $3.4\sigma$. Finally, the PAO has found evidence with a significance of $5.2\sigma$ of anisotropy in cosmic ray arrival directions at energies $>8\times10^{18}$ eV that is indicative of an extragalactic origin \citep{Auger2017b}. A recent analysis of correlations with potential sources yields significances of the excesses around Centaurus A and the most luminous active galactic nuclei (AGNs)  detected by \textit{Swift-BAT} of $\sim3\sigma$,  whereas for the star-forming galaxies there is a $4\sigma$ deviation from isotropy for energies greater than $39$ EeV at an intermediate angular scale of $13\deg$ \citep{Auger2017c,AugerStarburst2018}.
Besides these findings with cosmic ray observatories, the nearby starburst galaxies M82 and NGC 253 have been detected in gamma rays by the \textit{Fermi} satellite and Cherenkov telescopes \citep{acciari2009,acero2009,abdo2010}. The observations indicate that the cosmic ray density in the central regions of these galaxies, which is responsible for the $\gamma$-radiation, is hundreds of times that of our Galaxy. This overabundance of cosmic rays is expected because of the very high supernova rate in starbursts. The hadronic component of these rays should produce most of the gamma rays through $pp$ collisions with the ambient gas (e.g., \citealt{paglione1996,domingo2005,ohm2016}). The possible association of these gamma rays with the neutrino candidate events reported by IceCube in the energy range from $30$ to \mbox{$2000$ TeV} \citep{aartsen2014} is discussed by \citet{chang2014,chang2015} and \citet{anchordoqui2014}. Whether starbursts are actually responsible for the cosmic neutrinos detected by IceCube is still far from clear \citep{bechtol2017}.

Our goal in this paper is to make a new assessment, informed by the current astronomical data, of the potential of star-forming galaxies as cosmic-ray accelerators. The recent discoveries mentioned above make such a study timely. We shall focus on the extended superwind region; this region has been mostly neglected in previous discussions of cosmic and gamma ray production in starbursts. We shall adopt the southern starburst NGC 253 as a case study. Since the source is almost a twin galaxy of M82, our conclusions can be easily extrapolated to that source as well.  

The structure of the paper is as follows: in the next section we present the general picture of starburst-driven galactic superwinds as it emerges from the most recent observations. Then we characterize the superwind region of NGC 253 in Section \ref{NGC253}. Particle acceleration in the halo of this source is discussed in Section \ref{acceleration}. We analyze there both stochastic and diffusive shock accelerations. In Section \ref{losses} we evaluate the different losses and the maximum energies attainable. The temporal evolution of the particle distributions is presented in Section \ref{distributions}, whereas the corresponding spectral energy distributions (SEDs) of the photons produced by these particles are shown in Section \ref{SEDs}. In Section \ref{Discussion} we give a general discussion, including the potential contribution of a hidden supermassive black hole at the Galactic center. Finally, Section \ref{Conclusions} contains our conclusions. The Appendix is devoted to a discussion of the adiabaticity of the shocks in superwinds. 

\section{Superwinds from starburst galaxies}\label{Superwinds}

\begin{figure} 
\begin{center}
\includegraphics[trim= 0cm 0cm 0cm 0cm, clip=true, width=.47\textwidth,angle=0]{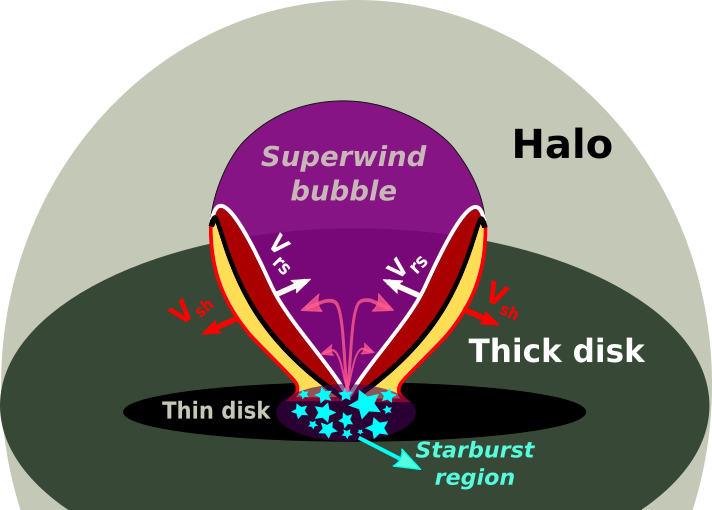}
\caption{Scheme of the physical scenario considered in this work (not to scale), adapted from \cite{strickland2002}.} 
\label{fig1}
\end{center}
\end{figure}

The basics of the superwind physics were laid out by \cite{chevalier1985} and developed by \cite{heckman1990} and \cite{strickland2002}, among others. The superwind occurs when the ejecta from supernovae and stellar winds in the nuclear region of the starburst is efficiently thermalized. The result is a very hot \mbox{($T\sim10^8$ K)} bubble with high pressure that expands and sweeps the ambient gas. When the bubble disrupts the disk it expands adiabatically into the halo, creating a multi-phased region. The outflow quickly reaches the terminal velocity given by 
\begin{equation}
  v_{\infty}\sim\sqrt{2\dot{E}/\dot{M}}\sim10^3\;\;\; \textrm{km\; s}^{-1}, \label{vinfty} 
\end{equation}
where $\dot{E}$ and $\dot{M}$ are the total energy release and the mass input. Simulations (e.g., \citealt{strickland2000}) show that the wind is bipolar and drags gas from the walls of the cavity. This shocked gas forms warm arcs \mbox{($T\sim10^4$ K)}. The swept material accumulated in front of the shocks forms a cold, dense shell around the bubble. The velocity of the shock in the dense medium is much slower than in the hot gas, reaching several hundreds of km s$^{-1}$. Figure \ref{fig1} shows a sketch of the situation, adapted from \cite{strickland2002}.

\subsection{Scaling relations}

Both $\dot{E}$ and $\dot{M}$ are proportional to the total contribution of supernovae and stellar winds in the central starbursts. If we denote by $*$ those quantities that correspond to these contributions, the total energy injected will be (e.g., \citealt{veilleux2005}, \citealt{tanner2017}):

\begin{equation}
  \dot{E}=\epsilon\; \dot{E}_*.
\end{equation}

Here, $\epsilon$ is the thermalization efficiency (the fraction of energy of the central SNe and stellar winds that goes into the outflow). This coefficient is highly dependent on the local conditions. Similarly, the total mass that goes into the outflow is formed by that supplied by the starbursts ($\dot{M}_*$) plus gas loaded by the wind from the ambient medium:

\begin{equation}
  \dot{M}=\dot{M}_* + \dot{M}_{\textrm{ISM}}=\beta\; \dot{M}_*.
\end{equation} 

The activity in the starburst is determined by the star formation rate (SFR). Both $\dot{E}_*$ and $\dot{M}_*$ scale with the SFR as \citep{veilleux2005}:

\begin{equation}
 \dot{E}_*= 7\times 10^{41} \; (\textrm{SFR}/ M_{\odot} \textrm{yr}^{-1})\;\; \textrm{erg s}^{-1}, \label{dotE}
\end{equation}

\begin{equation}
 \dot{M}_*= 0.26 \; (\textrm{SFR}/ M_{\odot} \textrm{yr}^{-1})\;\;M_{\odot} \textrm{yr}^{-1}. \label{dotM}\end{equation}

The supernova rate in the starbursts also scales with the SFR:

\begin{equation}
 \dot{\tau}_{\textrm{SN}}= 0.02 \; (\textrm{SFR}/ M_{\odot} \textrm{yr}^{-1})\;\; \textrm{yr}^{-1}.\label{tau}\end{equation}

Further, since the stars heat the dust that radiates in the infrared (IR), the SFR scales with the IR luminosity, which is a direct observable: 

\begin{equation}
  \textrm{SFR}\approx 17 \; \frac{L_{\textrm{IR}}}{10^{11} \textrm{erg}\; \textrm{s}^{-1} }\;\; M_{\odot}\; \textrm{yr}^{-1}.  \label{SFR}
\end{equation} 

It is worth noting that the terminal superwind velocity does not depend on the SFR:

\begin{equation}
  v_{\infty}= \sqrt{\frac{2\epsilon \dot{E}_*}{\beta \dot{M}_*}} \approx 3000 \sqrt{\frac{\epsilon}{\beta}}\; \textrm{km}\; \textrm{s}^{-1} .
\end{equation}  

The temperature in the central cavity of the starburst is:

\begin{equation}
  T= 0.4 \mu m_H \; \frac{\dot{E}}{k \; \dot{M}}\; \textrm{K},
  \end{equation}
where $k$ is Boltzmann constant, $m_H$ is the Hydrogen mass, and $\mu=1.4$.
Therefore, 
\begin{equation}
  T= 0.4 \; \mu \; m_H  \;\frac{\dot{E}}{k \; \dot{M}} \; \textrm{K} \approx 3 \times 10^8\; \frac{\epsilon}{\beta}\;\; \textrm{K}. \label{T}
 \end{equation}

The speed of the outer shell is smaller than the terminal velocity. In convenient units (e.g., \citealt{veilleux2005}):
\begin{equation}
  v_{\textrm{shell}}\approx 670\; \left(\frac{\epsilon \dot{E}_*}{n_{\rm s} \; 10^{44} \; \textrm{erg}\; \textrm{s}^{-1}}\right)^{1/3}\; R^{-2/3}_{\textrm{kpc}}  \; \textrm{km}\; \textrm{s}^{-1}.
\end{equation}  
In this expression $R_{\textrm{kpc}}$ is the radius of the superwind region in kpc and $n_{\rm s}$ the ambient particle density ahead of the shock in cm$^{-3}$. 

\subsection{Shocks and thermal cooling}

The bubble inflated by the superwind has a dynamical age of $t_{\textrm{dyn}}\sim 6 (v_{\infty} /1000 \;\textrm{km}\; \textrm{s}^{-1})^{-1}$ Myr. For a typical value \mbox{$v_{\infty}\sim 10^3$ km s$^{-1}$}, we get \mbox{$t_{\textrm{dyn}}\sim 6$ Myr}. The cooling time for thermal X-ray radiation of the plasma in the bubble is much longer ($t_{\textrm{cool}}\sim 77 \eta_{X}^{1/2} \cal{R}$$^{1/2}$ Myr, where the filling factor $\eta_{X}$ and the metallicity $\cal{R}$$=0.03/Z_{O, \; \textrm{true}}$ are parameters not very different from one\footnote{\cite{strickland2002} estimate, for the case of NGC 253, $Z_{O, \; \textrm{true}}\sim 0.5$, yielding $t_{\textrm{cool}}\sim 19$ Myr for a filling factor $\eta_{X}=1$.}). The bubble then expands freely and cools adiabatically, but it remains hot. For instance, in order to cool from \mbox{$3\times 10^6$ K} to  \mbox{$10^4$ K} a time \mbox{$t\sim 3\times 10^{5}\; (n/$cm$^{-3})^{-1}$ yr} is necessary. Typically  \mbox{$n\sim10^{-3}$ cm$^{-3}$} \citep{strickland2002}, meaning that more than \mbox{$10^8$ yr} are required. 

The expanding wind is surrounded by a boundary discontinuity that separates the hot gas from the swept-up shocked medium. A reverse shock moves through the free wind region. This shock is adiabatic and capable of efficient particle acceleration. The outer blast wave instead is radiative (see the Appendix). There, the dense gas cools down to warm temperatures  of $\sim 10^4$ K . This shell produces the H$\alpha$ emission associated with the superwind boundaries. The radiative shock is not suitable for efficient particle acceleration and is not considered below.  

Regarding the physical conditions in the hot wind region, the sound speed is 
\begin{equation}
  c_{\textrm{s}}=\sqrt{\frac{k T}{\mu m_H}}\approx 300 \left(\frac{T}{10^7 \; {\rm K}}\right)^{1/2}\;\; \textrm{km\; s}^{-1}.\label{sound}
   \end{equation} 

For typical values of $T$ inferred from X-ray observations, $c_{\textrm{s}}\sim 100$  km s$^{-1}$, and then the Mach number of the reverse shock is $\cal{M}$$\sim10$. Magnetic field values in this region, inferred from polarization radio observations, are of a few $\mu$G (e.g., \citealt{beck1994}). A  better characterization of the superwind region can be done for specific sources on the basis of multifrequency observations. In what follows we shall study the case of NGC 253, the most prominent starburst galaxy in the southern sky.

\section{The galactic-scale outflow of NGC 253}\label{NGC253}

NGC 253 is an edge-on starburst galaxy at a distance estimated in the range $2.6-3.9$ Mpc \citep{turner1985,puche1988,Karachentsev2003}. We shall adopt here a value of 2.6 Mpc as \cite{strickland2002}. NGC 253 is, along with M82, the best studied starburst galaxy. It has been observed and detected at all wavelengths, from radio to high-energy gamma rays. The existence of a superwind in this galaxy is well established. Diffuse X-ray emission with a temperature of $3\times10^6$ K from the halo has been detected by the \textit{Chandra} satellite \citep{strickland2002}. In the northern halo this emission seems to lie in the interior of a region delimited by a shell defined by the $H\alpha$ emission. Wind activity extends up to distances of \mbox{$\sim10$ kpc} from the nuclear starburst. The IR luminosity of the starburst is \mbox{$L_{\textrm IR}=1.7 \times 10^{10}$ $L_{\odot}$} \citep{ radovich2001}. Using Eq. (\ref{SFR}), the corresponding SFR is \mbox{SFR$_{\textrm{NGC}\;253}\sim 3$ $M_{\odot}$ yr$^{-1}$}. Using this value, Eqs. (\ref{dotE}-\ref{tau}) yield:

\begin{equation}
 \dot{E}= 2\times 10^{42}\; \epsilon \;\; \textrm{erg\;s}^{-1}, \label{dotE2}
\end{equation}
\begin{equation}
 \dot{M}= 0.75 \; \beta \; M_{\odot} \textrm{yr}^{-1} \label{dotM2},
 \end{equation}
and
\begin{equation}
 \dot{\tau}_{\textrm{SN}}= 0.058\;\; \textrm{yr}^{-1} \label{tau2},
 \end{equation}
 that is, one supernova every 17 years. 
 
 The mass load of the superwind of NGC 253 was recently determined through observations of $^{12}$CO $j=1-0$ transition lines obtained with the Atacama Large Millimeter Array (ALMA) performed by \cite{bolatto2013}. The wind is so heavy that it suppresses the star formation. A conservative estimate of the total molecular mass outflow rate is $\dot{M}\approx 9$ $M_{\odot}$ yr$^{-1}$. From this and Eq. (\ref{dotM2}) we get a mass loading factor $\beta\approx12$. The thermalization efficiency is constrained to the range 30\%--100\% in the case of M82 \citep{strickland2009}. Adopting a value of $\epsilon=0.75$, we obtain from Eqs. (\ref{dotE2}-\ref{dotM2}):
 \begin{equation}
 \dot{E}= 1.5 \times 10^{42}\; \;\; \textrm{erg\;s}^{-1}, \label{dotE3}
\end{equation}
\begin{equation}
 \dot{M}= 9 \; \; M_{\odot} \textrm{yr}^{-1} \label{dotM3},
 \end{equation}
 and, from Eq. (\ref{T}), a temperature for the starburst region of $T\sim 2\times 10^7$ K .
 
 The X-ray observations of the superwind region by \cite{strickland2002} allow us to determine the physical properties of the plasma in the inflated bubble. Modelling the region as a sphere of radius $R_{\textrm{kpc}}=5$ with an X-ray luminosity of $\sim 5\times 10^{38}$ erg s$^{-1}$ and a temperature $3\times 10^6$K, the sound speed is $c_{\textrm{s}}\approx164$ km s$^{-1}$, the particle density is $n_{\textrm{s}}\sim6.8\times 10^{-3}$ cm$^{-3}$ \citep{strickland2002}, the reverse shock velocity $v_{\textrm{rev}}\approx750$ km s$^{-1}$,  and the velocity of the expanding shell $v_{\textrm{shell}}\approx 298$  km s$^{-1}$. These latter values are only mildly changed if thermalization is highly efficient ($\epsilon\approx 1$): $v_{\textrm{rev}}\approx 866$ km s$^{-1}$ and $v_{\textrm{shell}}\approx 328$  km s$^{-1}$. 
  
 Multifrequency polarimetric radio observations of the halo of NGC 253 by \cite{heesen2009b} show polarized non-thermal emission that is identified with synchrotron radiation by cosmic ray electrons spiralling in the ambient magnetic field. \cite{heesen2009b} estimate the magnetic field in the halo in $B\sim 5$ $\mu$G. This implies a magnetic energy density of $\sim 1$ eV cm$^{-3}$, and with a volume of $\sim1.4\times10^{67}$ cm$^{3}$, a total magnetic energy of $\sim1.4\times10^{55}$ erg. The corresponding Alfv\'en velocity is 
 \begin{equation} 
v_{\textrm{A}}=\frac{B}{\sqrt{4\pi \rho}}\approx240\;\; \textrm{km\;s}^{-1}.  \label{va}
\end{equation}

 The radio observations by \cite{heesen2009} indicate that the cosmic ray transport from the disk to the northern halo is mainly convective, with a bulk velocity of $\sim 300$ km s$^{-1}$. The averaged diffusion coefficient, measured for the southern wind where the transport seems to be instead diffusive is $D\sim2 \times 10^{29}$ cm$^2$ s$^{-1}$.
 
 NGC 253 has been detected at gamma-rays by the \textit{Fermi} satellite in the GeV energy range \citep{abdo2010} and by the HESS telescope array at TeV energies \citep{acero2009}. The total emission at energies above 200 MeV corresponds to a luminosity of $\sim 4.3\times 10^{39}$ erg s$^{-1}$, when a distance of 2.6 Mpc is adopted (see \citealt{abramowski2012}). This radiation is thought to be produced mostly in the nuclear region \citep{domingo2005,rephaeli2010,abramowski2012}, although contributions from the superwind region cannot be ruled out.  
 
 In Table \ref{T1} we summarize the main parameters of NGC 253 and its superwind (SW) region. Table \ref{T2} shows the values that depend on the thermalization parameter $\epsilon$ for two different values (the mass load $\beta$ is fixed from observations to $\beta=12$).
 
 \begin{table}[ht]
    \caption[]{Physical properties of NGC 253 and its superwind.}
        \label{T1}
        \centering
\begin{tabular}{ll}
\hline\hline %
Starburst parameters & Value\\ [0.01cm]
\hline

$d$: Distance [Mpc]                                           &$2.6^1-3.9$\\
$L_{\rm{IR}}$:  Infrared luminosity [$L_{\odot}$]                                                                               & $1.7\times 10^{10}$\\
SFR:    Star-forming rate [$M_{\odot}$ yr$^{-1}$]                                                                & 3\\
$\beta$:      Mass loading factor                               & $12$\\
$\dot{M}$: Mass outflow [$M_{\odot}$ yr$^{-1}$] & $9$ \\
$L_{\gamma}(E>200\; \textrm{MeV})$:   $\gamma$-ray luminosity  [erg s$^{-1}$]                   & $4.3\times 10^{39}$\\
$T_\textrm{c}$: Temperature of the central region [K]$^2$ & $2\times 10^7$ \\

\hline\hline 
SW region parameters & Value \\ [0.01cm]
\hline
$R$:    Radius of the SW bubble [kpc]                                                                                   & $5$\\
$L_{\textrm x}$: X-ray luminosity [erg s$^{-1}$] & $5 \times 10^{38}$ \\
$T_{\textrm h}$: Temperature of the gas in the bubble [K] & $3\times10^6$\\
$n$: Particle density [cm$^{-3}$] & $2\times 10^{-3}$ \\
$B$: Magnetic field [$\mu$G] & 5 \\
$v_{\textrm A}$: Alfv\'en velocity [km s$^{-1}$] & 240\\
$v_{\textrm s}$: Sound speed [km s$^{-1}$] & 164\\
\hline
\hline  \\[0.005cm]
\end{tabular}                                                                           \tablefoot{$^1$ The value adopted in this paper. $^2$ Temperature estimated with Eq. (\ref{T}) adopting $\beta=12$ and $\alpha=0.75$.}                                                                                           
\end{table} 
 
  \begin{table*}
    \caption{Physical properties of the outflow depending on the thermalization parameter.}
        \label{T2}
        \centering
\begin{tabular}{lll}
\hline\hline %
Parameters & $\epsilon=1$ & $\epsilon=0.75$\\ [0.01cm]
\hline
$\dot{E}$: Mechanical luminosity of the superwind [erg s$^{-1}$] & $1.5 \times 10^{42}$ & $1.1 \times 10^{42}$\\
$v_{\textrm{rev}}$: Velocity of the reverse shock   [km s$^{-1}$]   &$866$ & $750$\\
$v_{\textrm{shell}}$: Velocity of the expanding shell   [km s$^{-1}$]   &$328$ & $298$\\
$c_{\textrm{s}}$: Sound speed in the hot central cavity [km s$^{-1}$]  &$474$ & $425$\\
\hline \\ [0.01cm]

\end{tabular}                                                                                                                                                                           
\end{table*}

\section{Particle acceleration}\label{acceleration}

As mentioned in the Introduction, the superwind regions of nearby starbursts are attractive potential sites for particle acceleration. We shall focus our discussion on the northern bubble of NGC 253, but our conclusions can be extended with some caution to M82. 

\subsection{General requirements}

The minimum requirement for a particle acceleration to an energy $E_{\textrm{max}}$ is that the magnetic field and the physical extent of the acceleration region are such that they can accommodate the particles. This is essentially the Hillas criterion \citep{hillas1984}. The Larmor radius (in cgs units) of a particle with energy $E$ is:
\begin{equation}
  r_{\textrm L}=\frac{E}{ZeB} \label{rL},
\end{equation}
where $Z$ is the atomic charge number, $e$ the elementary charge, and $B$ the magnetic field. If we assume that $r_{\textrm{L}}=R_{\textrm{kpc}}$, with $R_{\textrm{kpc}}$ the size scale of the halo created by the superwind in kpc, we can express the maximum theoretical energy as:
\begin{equation}
  E_{\textrm{max}}= 10^{18} Z R_{\textrm{kpc}} \left(\frac{B}{\mu\textrm{G}}\right) \;\; \textrm{eV} \label{Emax1}.
\end{equation}
From Table \ref{T1}, $R_{\textrm{kpc}}\sim 5$ and $B\sim 5$ $\mu$G. Then, we get:
\begin{eqnarray}
  E^p_{\textrm{max}}&=&2.5\times10^{19} \;\;\textrm{eV} \;\;\;\;\textrm{protons},\\ \label{Epmax1}
  E^{\textrm{Fe}}_{\textrm{max}}&=&6.5\times10^{20} \;\;\textrm{eV} \;\;\;\;\textrm{iron nuclei} \label{Efemax1}
  .\end{eqnarray}

This of course does not imply that the system will produce particles with such energies. It is only an absolute upper bound. Actual acceleration will depend on the efficiency of the acceleration process, the age of the accelerator, and the losses suffered by the particles. The result might be a much lower maximum energy, that might depend on several local factors.

Any real acceleration process will have an efficiency $\eta<1$. The timescale of the energy gain of the particles is given by  (e.g., \citealt{aharonian2002}):
\begin{equation}
  t_{\textrm{acc}}=E\left(\frac{dE}{dt}\right)^{-1}= r_{\textrm L}\frac{1}{\eta c}=\frac{E}{\eta ZeBc}.
  \end{equation} 
The energy gain can be written as :
\begin{equation}
  \dot{E}= \eta ZeBc.
  \end{equation} 
 The value of $\eta$ depends on the specific acceleration mechanism that operates in the source. In the case of the shocked plasma bubble created by the superwind in the halo of NGC 253 there are two possible acceleration mechanism that can result in the production of high-energy cosmic rays: stochastic diffusive acceleration (SDA) and diffusive shock acceleration (DSA) in the reverse adiabatic shock. We discuss them separately below. 

\subsection{Stochastic diffusive acceleration}

Since the plasma in the bubble created by the superwind is expected to be turbulent\footnote{The Reynolds number of the cosmic rays can be roughly calculated as \citep{bustard2017}: $$ {\cal R}=\frac{R_{\textrm{shock}} v_{\textrm{shock}}}{D}.$$ Adopting $D\sim 10^{29}$ cm$^2$ s$^{-1}$ \citep{heesen2009a}, we get ${\cal R}>10$.}, stochastic acceleration seems to be a viable mechanism for cosmic ray generation there. Observations of radio polarization suggest that the ratio of ordered to turbulent field is $\delta B/B_0\sim 1$ \citep{heesen2009b}. Particle interactions with hydromagnetic waves result in a net average energy gain that is of second order in the Alfv\'en velocity normalized to the speed of light (e.g., \citealt{ stawarz2008,o'sullivan2009,petrosian2012}). If the Alfv\'enic turbulence can be represented by a power spectrum of the form $W(k)\propto k^{-q}$, where $k$ is related to the wavelength $\lambda$ of the Alfv\'en waves by $k=2\pi/\lambda$, the turbulence can be written as \citep{o'sullivan2009}:
\begin{equation}
  \frac{\delta B^2}{8\pi}=\int^{k_{\textrm{max}}}_{k_{\textrm{min}}} W(k) \;dk,
\end{equation}
where $k_{\textrm{max}}$ and $k_{\textrm{min}}$ correspond to the shortest and longest wavelengths,. Adopting a spectrum with $q=1$, the acceleration timescale is \citep{stawarz2008,hardcastle2009}:

\begin{equation}
  t_{\textrm{acc}}=\left( \frac{v_{\textrm{A}}}{c} \right)^{-2} \left( \frac{\delta B^2}{B_0^2} \right)^{-1} \left(\frac{r_{\textrm{L}}}{c} \right).\label{tacc-SDA}
\end{equation}
In the case of the northern bubble of NGC 253, $\beta_{\textrm{A}}=v_{\textrm{A}}/c=8\times10^{-4}$, $\delta B^2/B_0^2\sim1$, and taking $r_{\textrm{L}}=5$ kpc, we get $t_{\textrm{acc}}\sim2.5\times10^{10}$ yr. Since this is of the order of the Hubble time, it is clear that the maximum theoretical energies cannot be reached. 

In absence of other losses, the maximum energy will be determined by the age of the source. The lifetime of the starbursts, and hence of the superwind activity, is much shorter than the age of the galaxy. An estimate of this age is given by the dynamical timescale, if the star-forming activity was approximately constant during the burst: $\tau\approx 2R_{\textrm{bubble}}/v_{\textrm{sw}}$, where  $2R_{\textrm{bubble}}\sim10$ kpc is the linear extent of the superwind region filled with the hot plasma and $v_{\textrm{sw}}$ is the expansion velocity of the gas in the IGM. For a thermal efficiency $\epsilon=0.75$, we get $\tau\approx10^7$ yr. Then, setting $t_{\textrm{acc}}=\tau$ we get the Larmor radius of the particles with the highest energies: $ r_{\textrm L}\sim 10^{19}$ cm. This corresponds to the following maximum energies:
\begin{eqnarray}
  E^p_{\textrm{max}}&=&1.5\times10^{16} \;\;\textrm{eV} \;\;\;\;\textrm{protons}  \label{Epmax2}\\ 
  E^{\textrm{Fe}}_{\textrm{max}}&=&4.0\times10^{17} \;\;\textrm{eV} \;\;\;\;\textrm{iron nuclei}. \label{Efemax2}
  \end{eqnarray}
These values are much smaller than those given by the Hillas's criterion. In the case of electrons, radiative losses make the maximum energies significantly lower than for protons (see Sect. \ref{losses} below). 

The total luminosity in cosmic rays is:
\begin{equation}
 L_{\textrm{CR}}\sim \zeta \frac{B^2}{8\pi} v_{\textrm{A}} (4\pi R_{\textrm{bubble}}^2),  
\end{equation}
where  $\zeta$ is the fraction of magnetic energy that is converted into cosmic rays. For $\zeta\sim 10$ \% we get:
\begin{equation}
 L_{\textrm{CR}}\sim 6.7 \times 10^{39}\;\;\; \textrm{erg\;s}^{-1}. 
 \label{LcrDSA}
\end{equation}

\subsection{Diffusive shock acceleration}

The reverse shock created by the interaction of the superwind with the external medium is adiabatic and hence suitable in principle to accelerate particles by a faster Fermi type I mechanism. If the shock is strong and super-Alfv\'enic, in the test particle approximation, the energy gain (in cgs units) is (e.g.,   
 \citealt{drury1983,protheroe1999}):
\begin{equation}
  \frac{dE}{dt}=\frac{3}{20} e c \left( \frac{D}{D_{\textrm{B}}}\right)^{-1} \left( \frac{v_{\textrm{rev}}}{c}\right)^{2} B,
\end{equation}
where $D$ is the diffusion coefficient at the shock in Bohm units: $D_{\textrm{B}}=cr_{\textrm{L}}/3$. The acceleration timescale is:

\begin{equation}
  t_{\textrm{acc}}\approx 2.1 \left( \frac{D}{D_{\textrm{B}}}\right) \left( \frac{v_{\textrm{rev}}}{1000\;\textrm{km}\;\textrm{s}^{-1}} \right)^{-2} \left( \frac{B}{\mu\textrm{G}}\right)^{-1} \left(\frac{E}{\textrm{GeV}}\right)\;\; \textrm{yr}.
\end{equation}
Then, equating this to the lifetime of the source, $ t_{\textrm{acc}}=\tau$, we get in the Bohm limit $D=D_{\textrm{B}}$ and for a thermalization $\epsilon=1$, the following maximum particle energies for the parameters of NGC 253:
\begin{eqnarray}
  E^p_{\textrm{max}}&=&1.7\times10^{16} \;\;\textrm{eV} \;\;\;\;\textrm{protons}  \label{Epmax3}\\ 
  E^{\textrm{Fe}}_{\textrm{max}}&=&4.4\times10^{17} \;\;\textrm{eV} \;\;\;\;\textrm{iron nuclei}. \label{Efemax3}
  \end{eqnarray}
These values are quite similar to those obtained for the case of SDA. For $\epsilon=0.75$ ($v_{\textrm{rev}}=750$ km s$^{-1}$), the values reduce to $1.2\times 10^{16}$ and $3.1\times10^{17}$ eV, respectively. 

\subsubsection{Shock luminosity}\label{sh-lum}

The cosmic ray luminosity produced by DSA at the terminal shock is:
\begin{equation}
  L_{\textrm{CR}}=4\pi \xi R^2_{\textrm{bubble}} \rho v^3_{\textrm{shock}}\sim \xi \dot{M} v^2_{\textrm{shock}}.
   \end{equation}
Here, $\xi$ is the efficiency of converting shock kinetic energy into cosmic rays. For $v_{\textrm{shock}}\sim v_{\textrm{rev}}\sim750$ km s$^{-1}$ and $\xi\sim0.1$, we get: $L_{\textrm{CR}}\sim 3.2 \times 10^{41}$ erg s$^{-1}$, which is more than an order of magnitude above the cosmic ray power generated in our Galaxy. The actual value of $\xi$ can be constrained by astronomical observations of the non-thermal radiation produced by particles. We discuss these constraints in Section \ref{SEDs}.

\section{Losses and highest energies}\label{losses}

\begin{figure} 
\begin{center}
\includegraphics[trim= 0cm 0cm 0cm 0cm, clip=true, width=.47\textwidth,angle=0]{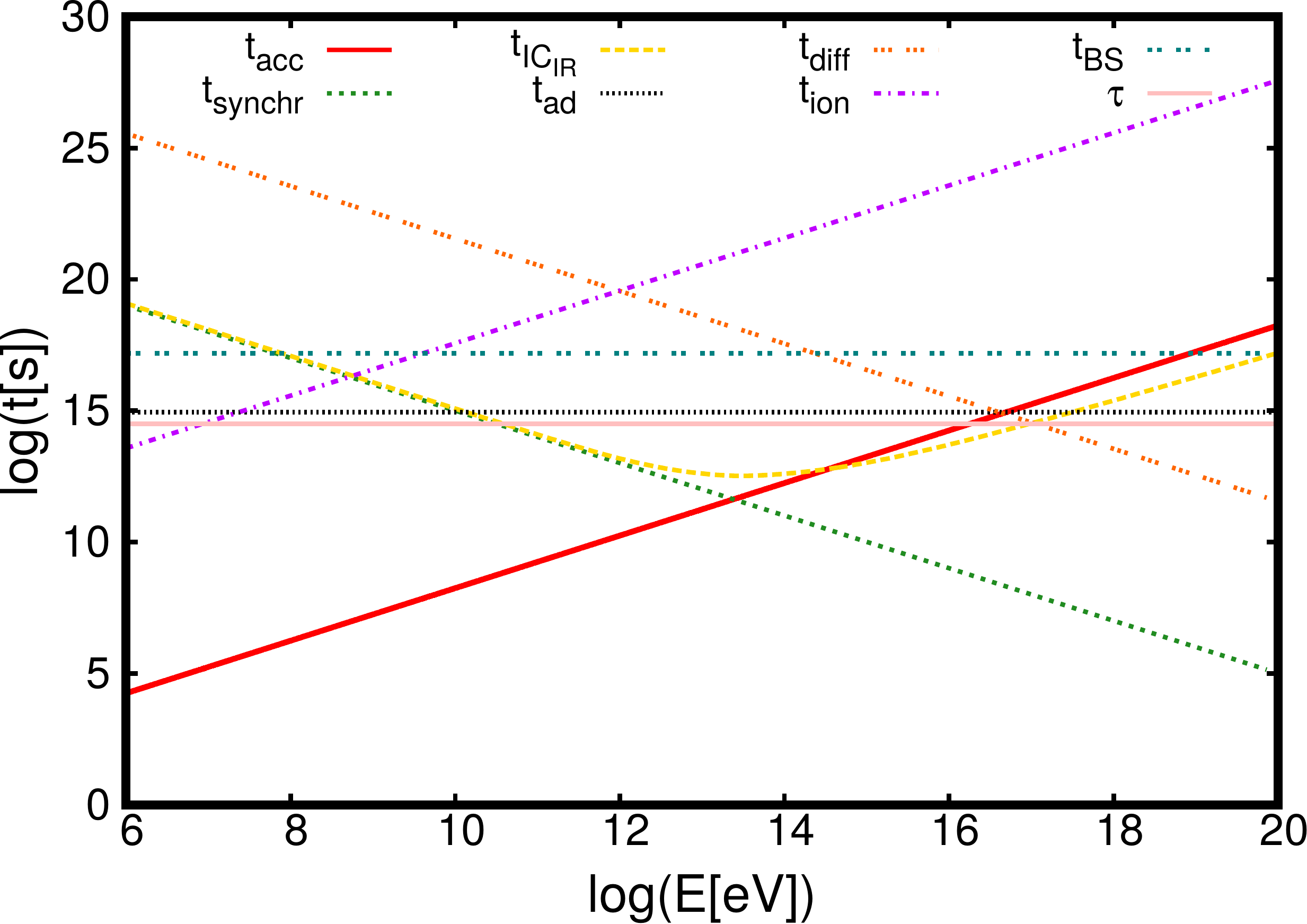}
\caption{Acceleration and cooling times for the electrons in a $5\mu$G magnetic field with a thermalization $\epsilon=1$.} 
\label{t_e_term1_5uG}
\end{center}
\end{figure}

In order to obtain final estimates for the highest energies of cosmic rays accelerated in the superwind region of NGC 253 we need to ponder the losses that different species of particles can undergo along the accelerations processes. These losses depend on the conditions in the source and can be divided into radiative and non-radiative losses. For electrons they are caused by synchrotron radiation, inverse Compton (IC) up-scattering of low-energy photons, and relativistic Bremsstrahlung (BS). The non-radiative losses are due to ionization and adiabatic cooling. Below we\ provide suitable expressions for the timescales of all these losses (see, e.g., \citealt{bosch-ramon2010} and references therein):

\begin{eqnarray}
  t_{\textrm{syn}}&\sim& 8.3 \times10^9 \; \left(\frac{E}{\textrm{GeV}}\right)^{-1} \left(\frac{B}{\mu\textrm{G}}\right)^{-2} \;\; \textrm{yr}, \label{syn}\\ 
  t_{\textrm{BS}}&\sim&3.9 \times10^7 \; \left(\frac{n}{\textrm{cm}^{-3}}\right)^{-1}  \;\; \textrm{yr}, \label{brems}\\  
  t_{\textrm{ion}}&\sim& 9.5 \times10^6 \; \left(\frac{n}{\textrm{cm}^{-3}}\right)^{-1} \left(\frac{E}{\textrm{GeV}}\right) \;\; \textrm{yr}, \label{ion}\\ 
  t_{\textrm{ad}}&\sim& 3  \; \left(\frac{dv_{\textrm{shock}}}{dz}\right)^{-1}\nonumber \\ &\sim & 4.8 \times 10^6  \left( \frac{v_{\textrm{shock}}}{1000\;\textrm{km}\;\textrm{s}^{-1}} \right)^{-1} R_{\textrm{kpc}}  \;\; \textrm{yr}. \label{ad}   
   \end{eqnarray}
Specifically, we use the expression provided by \citet{khangulyan2014} for the calculation of the IC cooling time.
Adopting the values of NGC 253, we find that the main losses are caused by synchrotron radiation and IC scattering. The photon fields in the superwind region are the cosmic microwave background with $U_{\textrm{ph}}\sim 0.25$ eV cm$^{-3}$ and the IR field, which is the dominant one, with $U_{\textrm{ph}}\sim 1$ eV cm$^{-3}$. Using the latter value and  $B\sim5\mu$G,  we see that $ t_{\textrm{syn}}\sim  t_{\textrm{IC}}$ (see Fig. \ref{t_e_term1_5uG}). Subsequently, equating $t_{\textrm{acc}}^{-1}= t_{\textrm{syn}}^{-1}+ t_{\textrm{IC}}^{-1}\sim 2/t_{\textrm{syn}}$, we find for DSA that $E^e_{\textrm{max}}\approx 35$ TeV ($\epsilon=1$) and 30 TeV ($\epsilon=0.75$).


Stochastic acceleration is a slower process. For the Alfv\'en velocity estimated for NGC 253, Eq. (\ref{tacc-SDA}) can be rewritten as:
\begin{equation}
  t_{\textrm{acc}}=5.5 \left( \frac{E}{\textrm{GeV}} \right) \left( \frac{B}{\mu\textrm{G}} \right)^{-1} \textrm{yr}.\label{tacc-SDA2}
\end{equation}
The maximum energies achieved by electrons are lower than in the case of DSA:  $E^e_{\textrm{max}}\approx 17$ TeV.

\begin{figure} 
\begin{center}
\includegraphics[trim= 0cm 0cm 0cm 0cm, clip=true, width=.47\textwidth,angle=0]{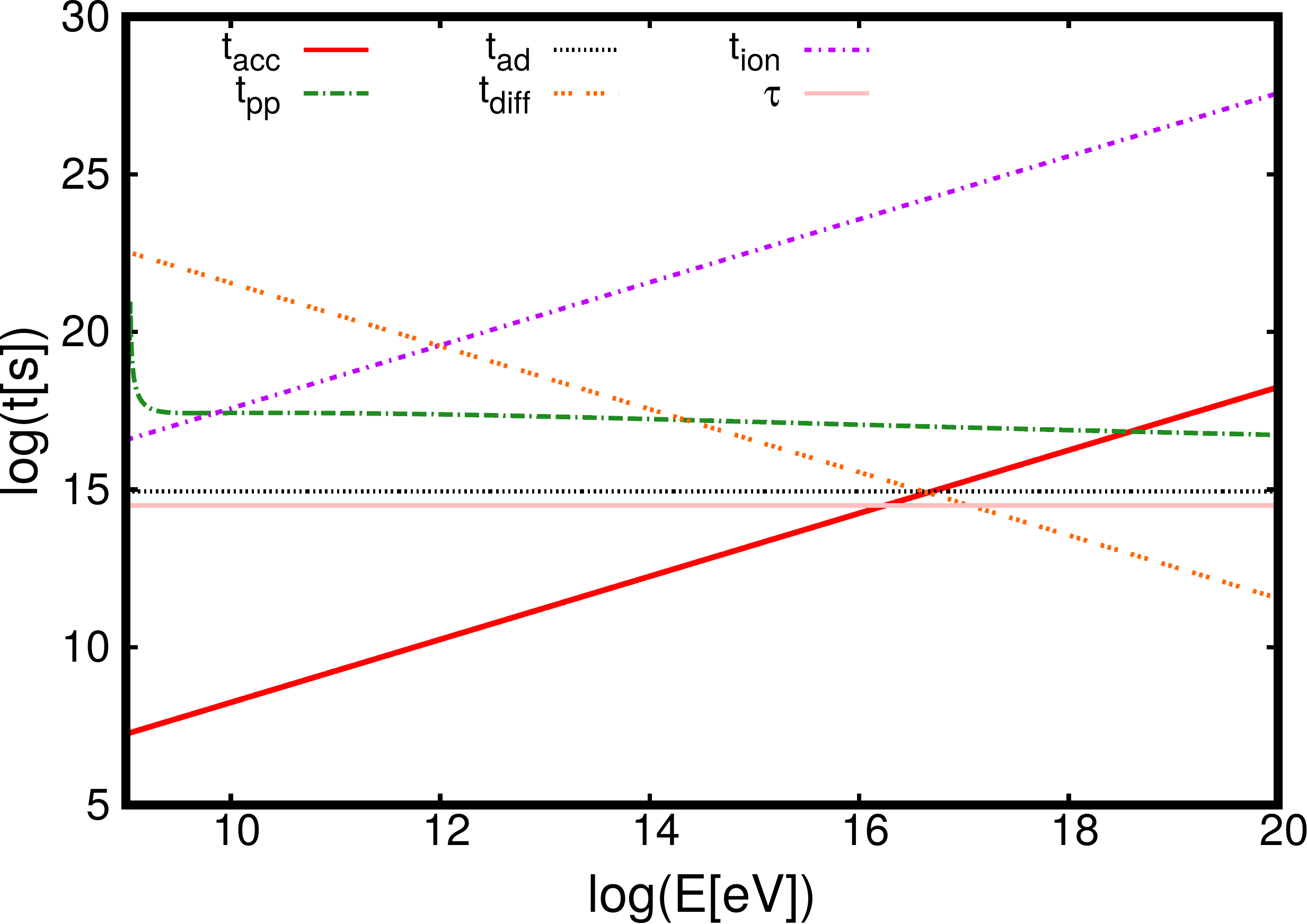}
\caption{Acceleration and cooling times for the protons in a $5\mu$G magnetic field with a thermalization $\epsilon=1$.} 
\label{t_p_term1_5uG}
\end{center}
\end{figure}

Protons are affected by $pp$ interactions, photo-pion production, adiabatic losses, and escape by diffusion. Under the physical conditions in the superwind region, all timescales associated with these processes are longer than the dynamical age of the region, so our previous estimates of maximum energies are not affected. The results are shown in Fig. \ref{t_p_term1_5uG}, for a thermalization parameter $\epsilon=1$. In the calculations illustrated by this plot we used the formulae and approximations given by \cite{kelner2006}. Diffusive escape is calculated in the Bohm limit. 

\section{Particle distributions}\label{distributions}

A determination of the spectrum $N_i(E, t)$ of the particles accelerated in the superwind region of the starbursts requires solving the transport equation:
\begin{eqnarray}
  \frac{\partial N_{e,p}(E, t)}{\partial t}&=& \frac{\partial}{\partial E} \left[ D(E) \frac{\partial N_{e,p}(E, t)}{\partial E}\right] \nonumber \\
  && - \frac{\partial}{\partial E} \left[ \left(\frac{2D(E)}{E}  - b_{e,p}(E)\right)  N_{e,p}(E, t) \right] \nonumber \\ && + \dot{N}_{e,p;\;\textrm{inj}}. \label{Teq1}
\end{eqnarray}
Here, the index $e, p$ indicates electrons and protons, respectively. The term $b(E)=\dot{E}$ represents the different losses. For protons in the present situation this term can be set to zero, and for electrons it should contain both synchrotron and IC losses. $D(E)$ is the energy diffusion coefficient in the region with magnetic turbulence.



The mechanism that excites the turbulence is not known. A general energy diffusion coefficient can be defined as (e.g., \citealt{asano2016}):
\begin{equation}
  D(E)=K E^{q}. \label{dif}
\end{equation}
For simplicity we shall adopt the most optimistic case $D(E)\propto E$ that is the counterpart of the Bohm limit in DSA. An index $q=1$ is observed, for instance, when Alfv\'en waves are the scatterers in magnetohydrodynamic simulations by  \cite{kowal2010}. The parameter $K$ in Eq. (\ref{dif}) can by approximated as $K\sim t_{\textrm{acc}}^{-1}$. Subsequently, 
\begin{equation}
  D(E)=7.4 \times 10^{-14} \left(\frac{E}{\textrm{GeV}}\right)\;\;\; {\rm erg^2\;s^{-1}}.\label{dif2}
\end{equation}

\begin{figure} 
\begin{center}
\includegraphics[trim= 0cm 0cm 0cm 0cm, clip=true, width=.47\textwidth,angle=0]{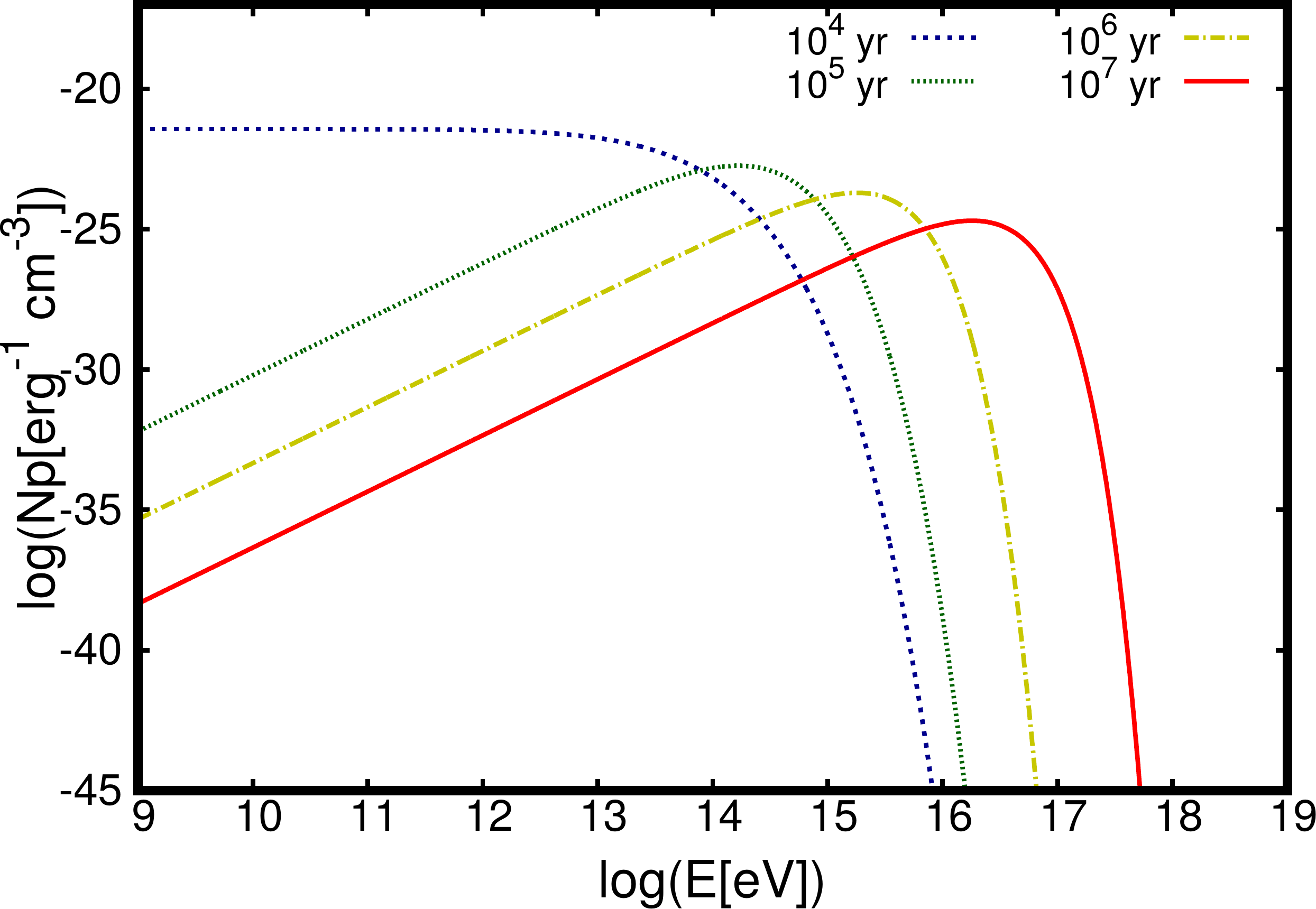}
\caption{Proton distributions for $a=1$ with $a=L_{\rm p}/L_{\rm e}$ in the case of SDA for different times.} 
\label{Np_dsa_a1}
\end{center}
\end{figure}

\begin{figure} 
\begin{center}
\includegraphics[trim= 0cm 0cm 0cm 0cm, clip=true, width=.47\textwidth,angle=0]{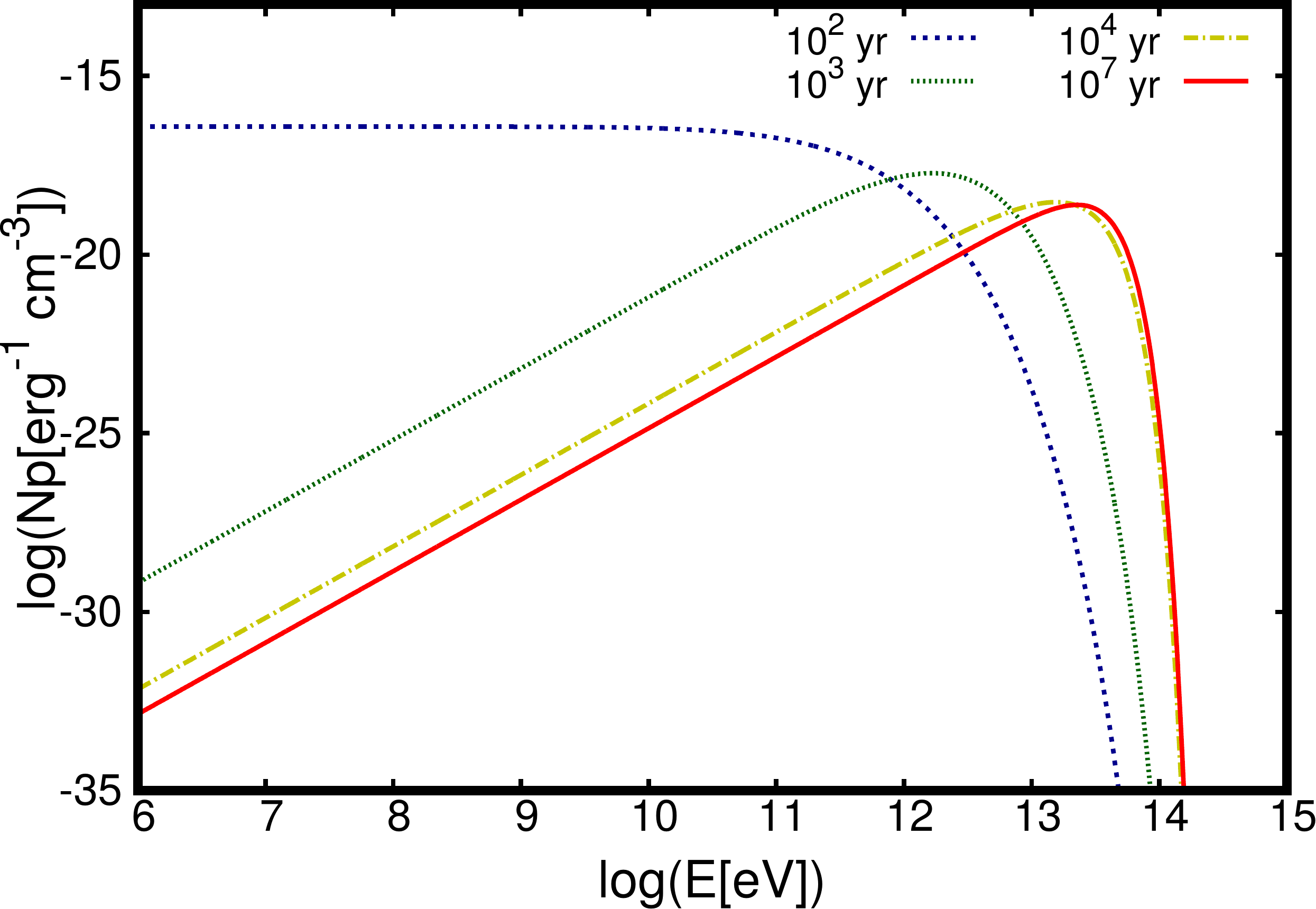}
\caption{Electron distributions for $a=1$ with $a=L_{\rm p}/L_{\rm e}$ in the case of SDA for different times and the final steady state solution.} 
\label{Ne_dsa_a1}
\end{center}
\end{figure}

For pure SDA of protons (no losses), Eq. (\ref{Teq1}) reduces to (e.g., \citealt{o'sullivan2009,chang1970}):
\begin{eqnarray}
  \frac{\partial N_{p}(E, t)}{\partial t} &=& \frac{\partial}{\partial E} \left[ D(E) \frac{\partial N_{p}(E, t)}{\partial E}\right]\\ &-&\frac{\partial}{\partial E} \left[\frac{2D(E)}{E}  N_{e,p}(E, t)\right]. \label{Teq2} 
\end{eqnarray}
This equation can be solved for the diffusion coefficient given in Eq. (\ref{dif2}) using numerical methods (e.g., \citealt{park1996}). In Fig.  \ref{Np_dsa_a1} we show the proton  distribution for different times. Because of the importance of the IC and synchrotron radiative losses and the slow acceleration rate in SDA,  the electrons reach the steady state after \mbox{$10^{4}$ yr}. Figure \ref{Ne_dsa_a1} shows the solution of Eq. (\ref{Teq1}) for electrons at different times until they became steady. We adopt in our calculations the Maxwell-Boltzmann energy distribution as initial particle
distribution with $T=3\times10^{6}$ K and impose the constraint that the available energy is limited by Eq. \ref{LcrDSA}.
 

\begin{figure} 
\begin{center}
\includegraphics[trim= 0cm 0cm 0cm 0cm, clip=true, width=.47\textwidth,angle=0]{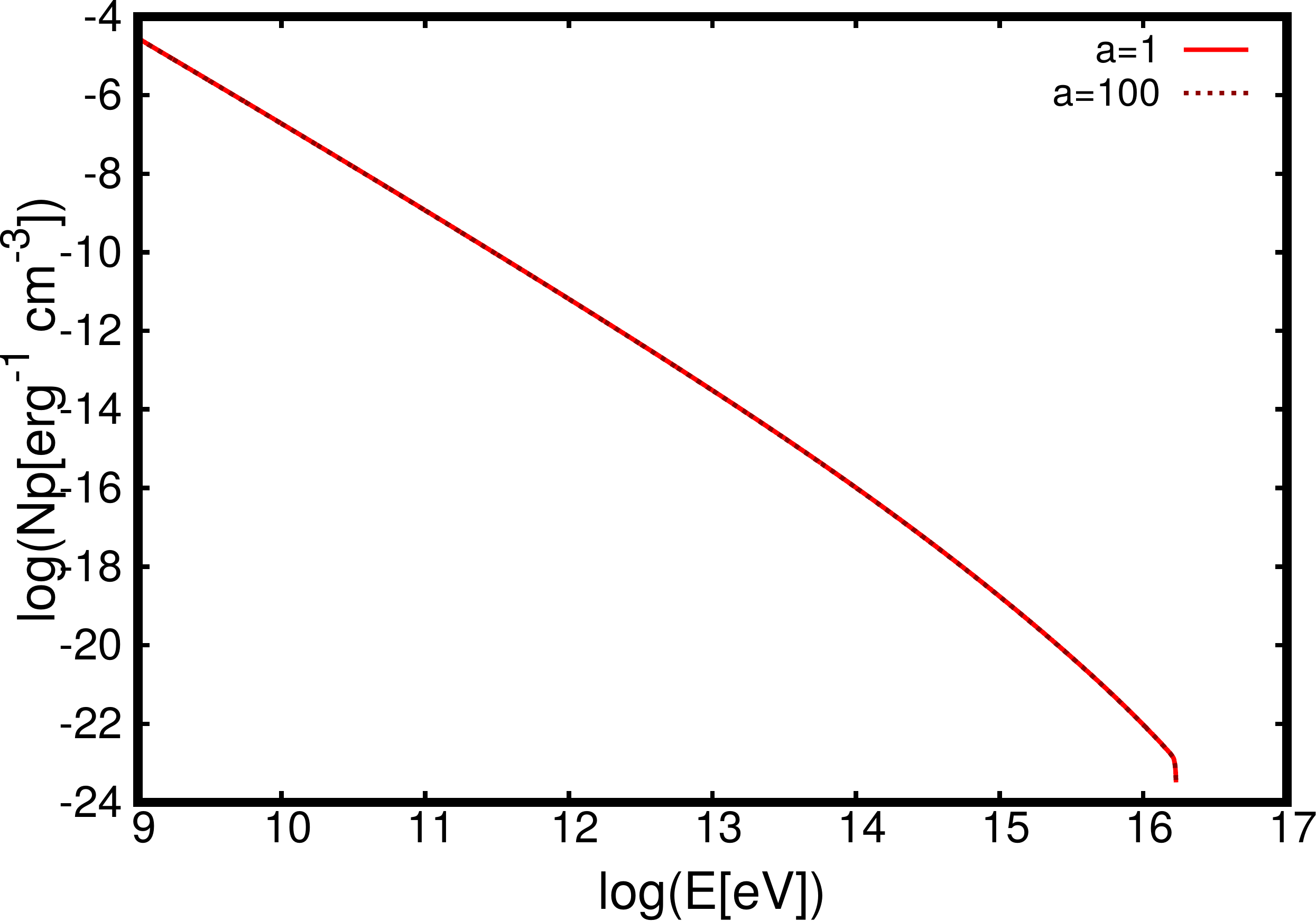}
\caption{Proton distributions for $a=1$ and $a=100$ with $a=L_{\rm p}/L_{\rm e}$ in the case of DSA in the reverse shock, assuming thermalization $\epsilon=1$.} 
\label{proton_distrib}
\end{center}
\end{figure}

\begin{figure} 
\begin{center}
\includegraphics[trim= 0cm 0cm 0cm 0cm, clip=true, width=.47\textwidth,angle=0]{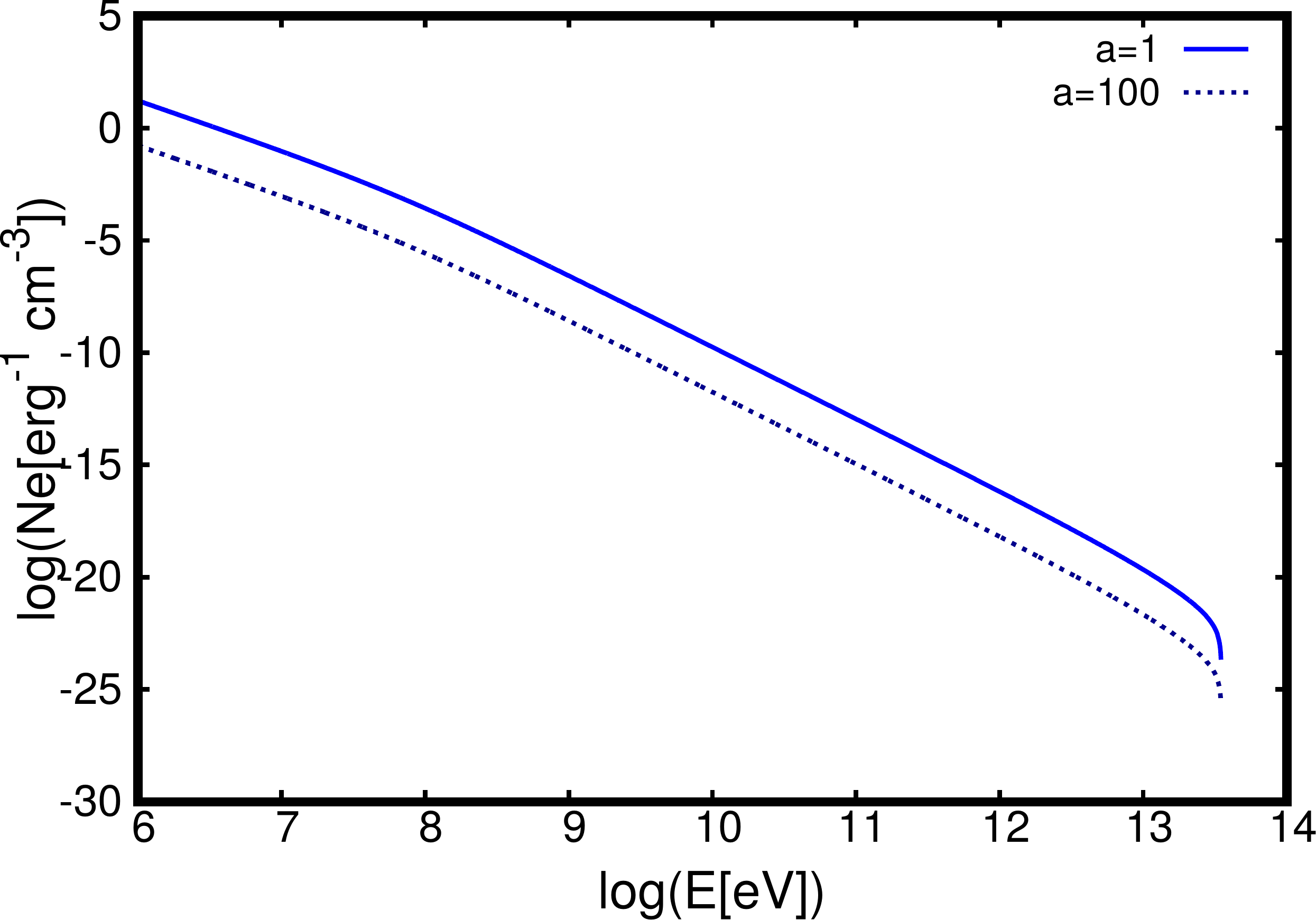}
\caption{Electron distributions for $a=1$ and $a=100$ with $a=L_{\rm p}/L_{\rm e}$ in the case of DSA in the reverse shock, assuming thermalization $\epsilon=1$.} 
\label{electron_distrib}
\end{center}
\end{figure}

Finally, Figs. \ref{proton_distrib} and \ref{electron_distrib} present solutions for the transport equation corresponding to the case of DSA with a constant power-law injection of index $-2.2$.  This equation can be written as (e.g., \citealt{ginzburg}):
\begin{equation}
  \frac{\partial N_{e,p}(E, t)}{\partial t}+\frac{\partial[b(E) N_{e,p}(E,t)]}{\partial E}+ \frac{N_{p}(E, t)}{t_{\textrm{esc}}}=Q(E). \label{Teq3}
\end{equation}
Here $Q(E)$ is the injection term and $t_{\textrm{esc}}$ is the escape time, which is dominated by the spatial diffusion time $t_{\textrm{diff}}=R^2/D(E)$. Only particles with $t_{\textrm{diff}}$ shorter than the dominant cooling timescale can escape from the accelerator into the IGM. We consider Bohm diffusion at the shock. For a magnetic field of 5 $\mu$G this means \mbox{$D_{\textrm{B}}(E)= 6.6\times10^{21} (E/\textrm{GeV}) Z^{-1} $ cm$^2$ s$^{-1}$}.

We consider two cases: one where the power in protons and electrons is the same and another where protons dominate by two orders of magnitude, as in the cosmic rays of our Galaxy. If we define $a=L_p/L_e$, with $L_i$ the power in relativistic particles of species $i$, these cases correspond to $a=1$ and $a=100$, respectively. 

\section{Spectral energy distributions}\label{SEDs}

\begin{figure} 
\begin{center}
\includegraphics[trim= 0cm 0cm 0cm 0cm, clip=true, width=.47\textwidth,angle=0]{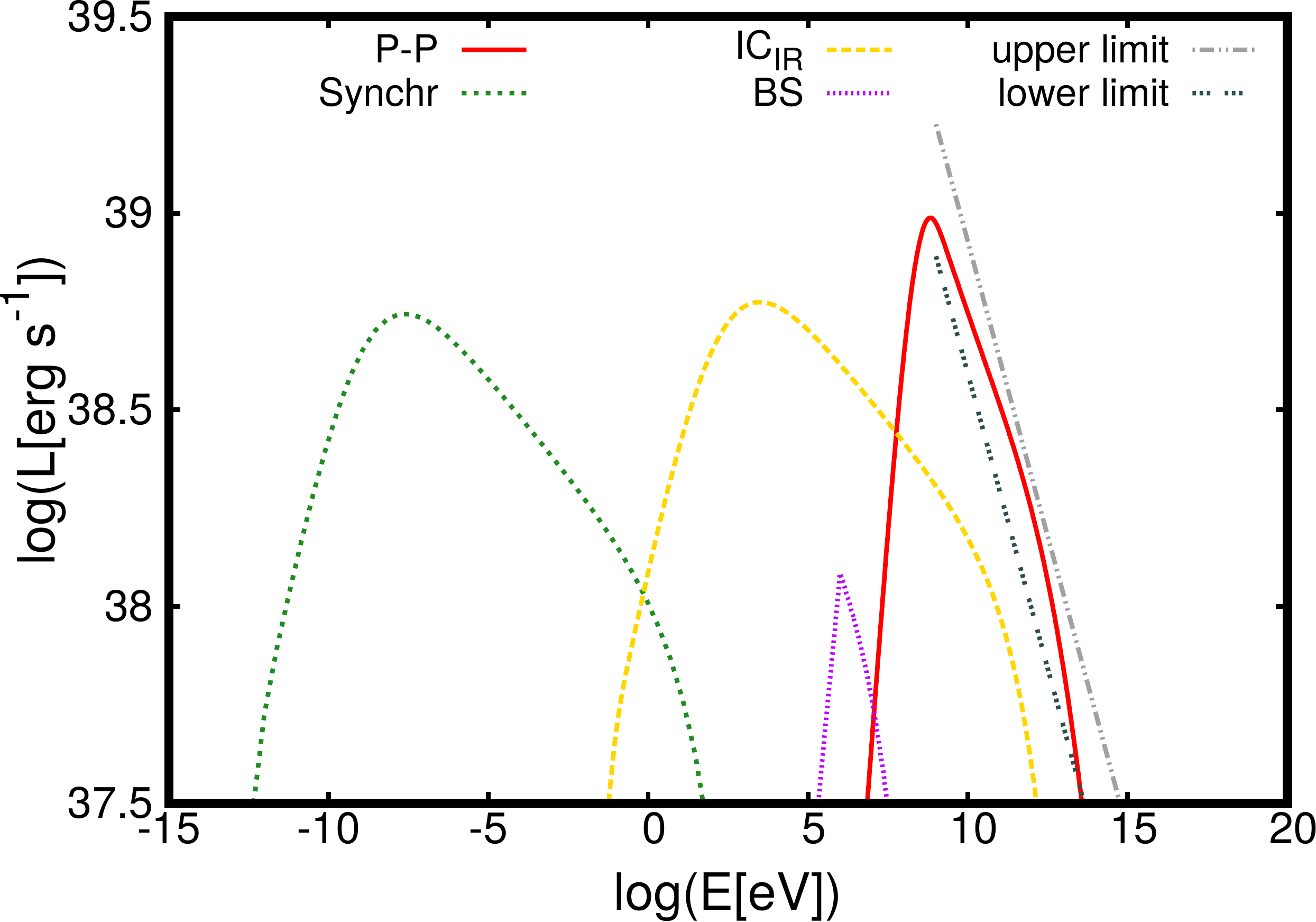}
\caption{Spectral energy distribution in the case of DSA with ratio $a=1$, magnetic field $B=5\mu$G, thermalization $\epsilon=1$, and shock efficiency $\xi=0.012$.} 
\label{SED_term1_5uG_a1}
\end{center}
\end{figure}

\begin{figure} 
\begin{center}
\includegraphics[trim= 0cm 0cm 0cm 0cm, clip=true, width=.47\textwidth,angle=0]{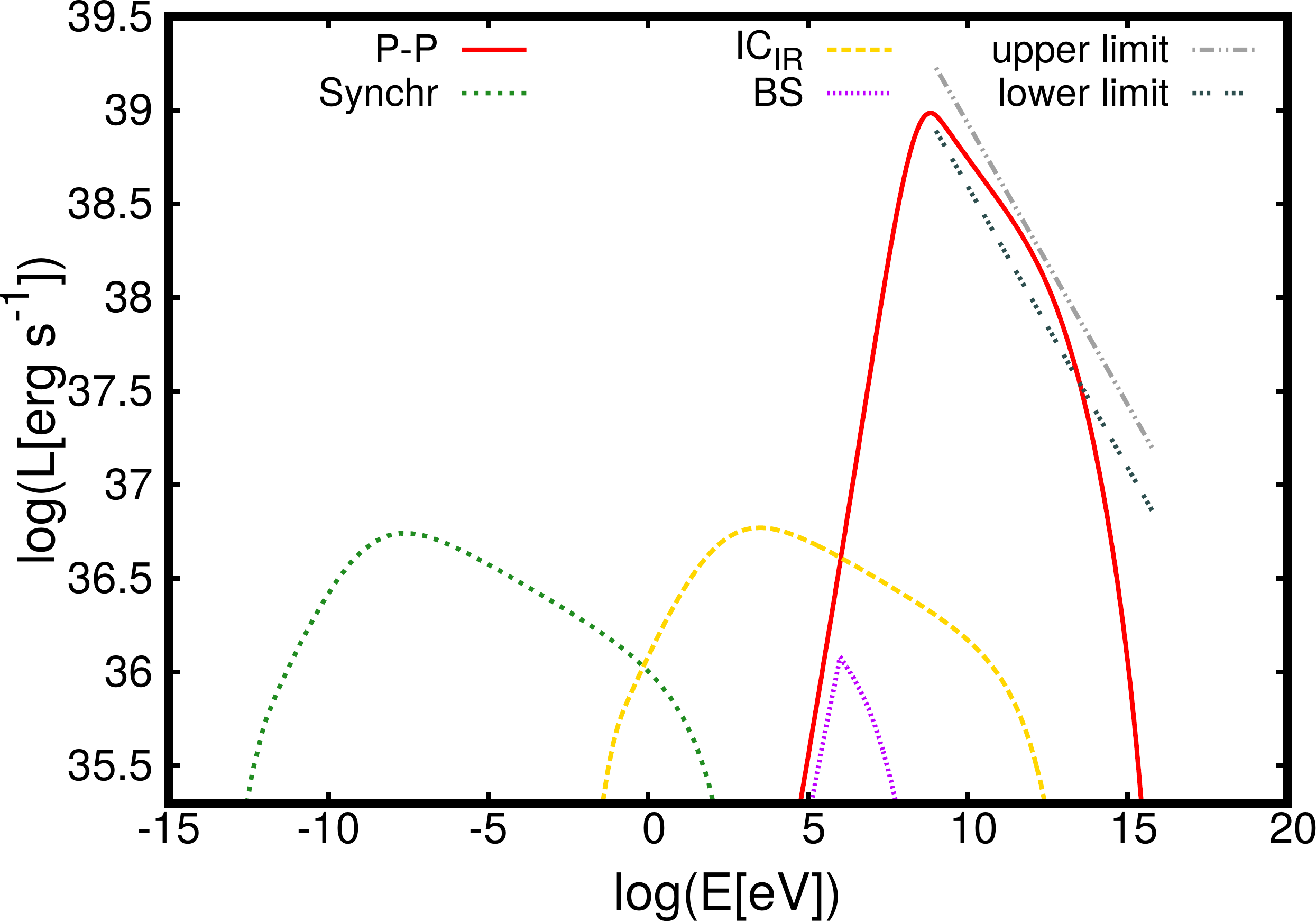}
\caption{Spectral energy distribution in the case of DSA with ratio $a=100$, magnetic field $B=5\mu$G,  thermalization $\epsilon=1$, and shock efficiency $\xi=6 \times 10^{-3}$.} 
\label{SED_term1_5uG_a100}
\end{center}
\end{figure}

In this section we present the SEDs of the radiation produced by the cosmic rays in the halo of NGC 253. Different acceleration mechanisms predict different maximum energies for electrons. Hence the radiation they produce, and in particular the location of the cut off of their emission, can be used to discriminate between DSA and SDA in the superwind region. Figure \ref{SED_term1_5uG_a1} shows the SED obtained in the case of DSA and ratio $a=1$, whereas Fig. \ref{SED_term1_5uG_a100} shows the case of $a=100$.  We have used the expressions provided by \citet{vilaaharonian2009} to calculate the emissions. The efficiency of the shock to convert kinetic energy into cosmic rays, $\xi$, has been adjusted for the SEDs in gamma-rays not to exceed the limits imposed for \textit{Fermi} and HESS (e.g., \citealt{lacki2011}). We see that only $\sim$1\% of the shock power should go to cosmic rays. 

\begin{figure} 
\begin{center}
\includegraphics[trim= 0cm 0cm 0cm 0cm, clip=true, width=.47\textwidth,angle=0]{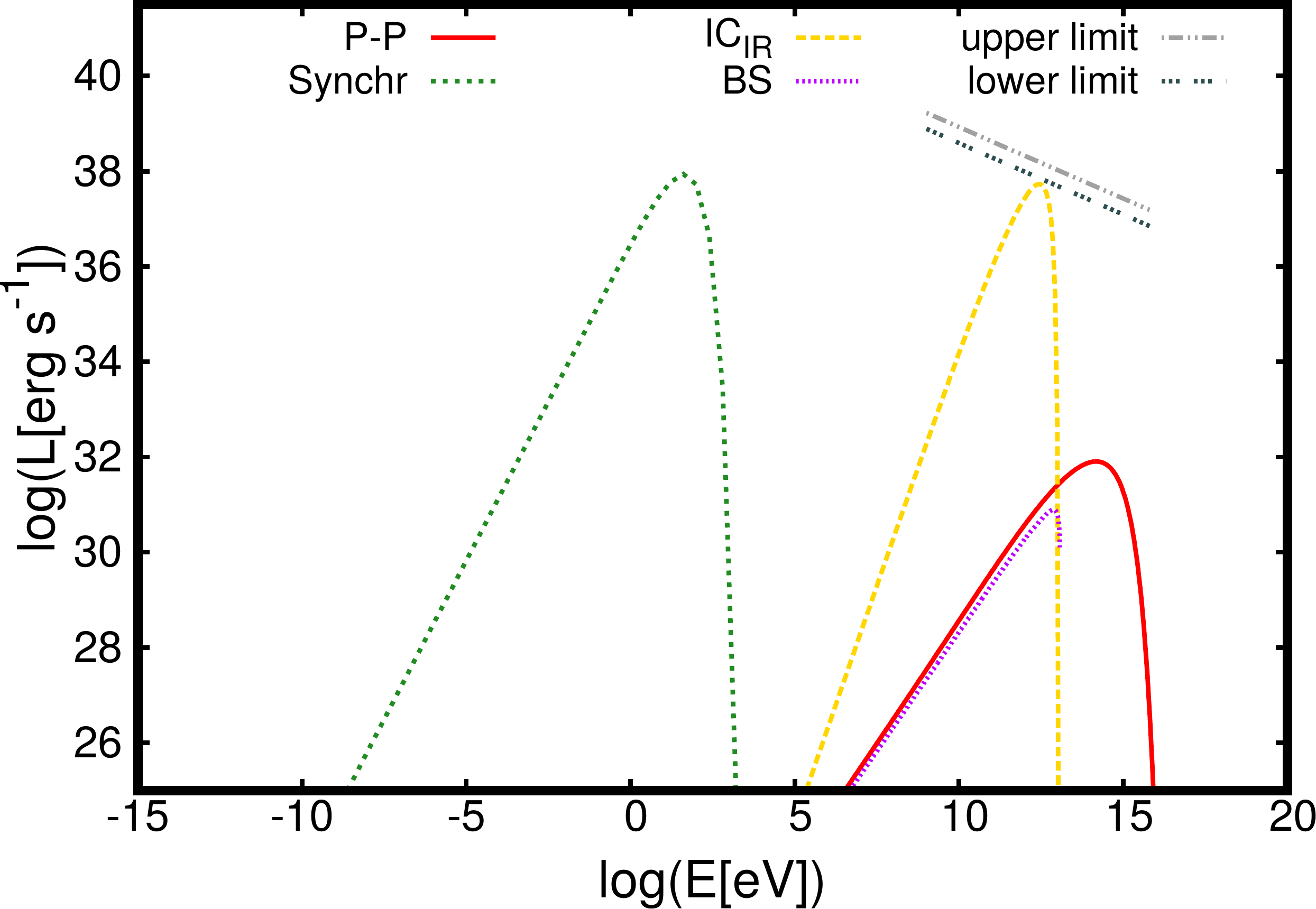}
\caption{Spectral energy distribution in the case of SDA with ratio $a=1$, magnetic field $B=5\mu$G.} 
\label{SED_dsa_a1}
\end{center}
\end{figure}

\begin{figure} 
\begin{center}
\includegraphics[trim= 0cm 0cm 0cm 0cm, clip=true, width=.47\textwidth,angle=0]{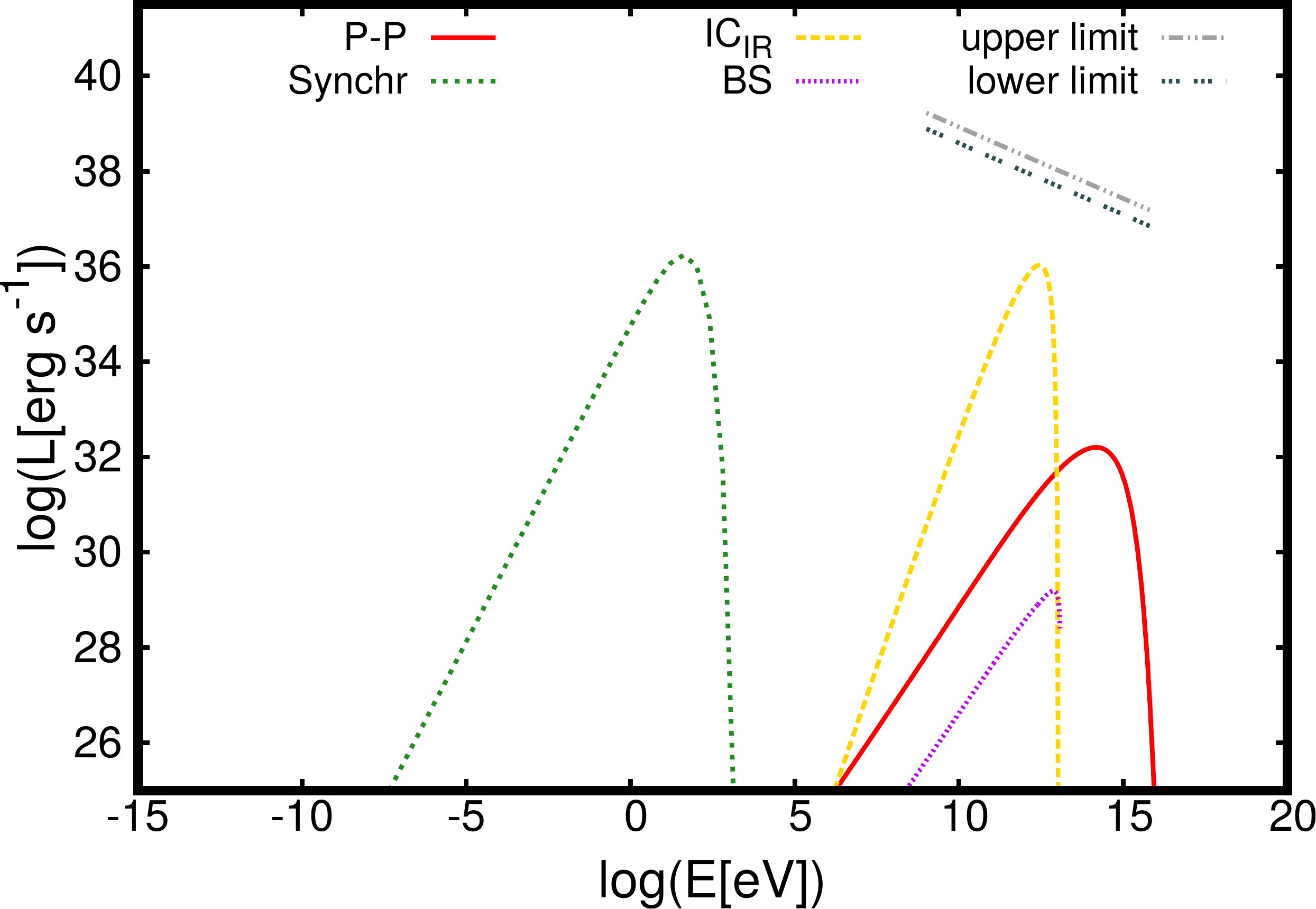}
\caption{Spectral energy distribution in the case of SDA with ratio $a=100$, magnetic field $B=5\mu$G.} 
\label{SED_dsa_a100}
\end{center}
\end{figure}

Figures \ref{SED_dsa_a1} and \ref{SED_dsa_a100} present the SEDs corresponding to SDA, for the same ratios. 

In principle, X-ray observations can be used to differentiate the models, since in the case of DSA non-thermal X-rays are present (see Fig. \ref{SED_term1_5uG_a1}).

\subsection{Comparison with other starbursts}\label{other}

In addition to NGC 253, the nearby starburst galaxies M 82, NGC 4945, and NGC 1068 have been detected by the \textit{Fermi} satellite and Imaging Atmospheric Cherenkov Telescopes (e.g., \citealt{Ackermann2012}, \citealt{Fermi3rdCatalogue}). The main characteristics of the gamma-ray emission of these sources are given in Table \ref{tab:Lcd}. Given the general similarities between NGC 253 and M 82, it is not surprising that these two sources are also quite alike at high energies. They both present spectra with indices of around $-2.2$, which are suggestive of DSA (SDA tends to produce harder spectra, see Figs. \ref{Np_dsa_a1}-\ref{electron_distrib}). However, it is important to emphasize that none of these sources is resolved in gamma-rays, so the observed emission might be dominated by contributions from the disk (e.g., \citealt{Bykov2014}). 

The gamma radiation of the galaxy NGC 4945 is also very similar. Its central region, however, harbors a supermassive black hole of $\sim10^6$ $M_{\odot}$ in addition to the star-forming region. Recent investigations of the correlation between the gamma-ray and the X-ray emission strongly support the theory that the origin of the high-energy radiation is related to the AGN activity, instead of the starburst \citep{NGC4945}.

Finally, NGC 1068 is the nearest Seyfert 2 galaxy. The gamma-ray luminosity is more than one order of magnitude higher than in the case of NGC 253. Most of this power seems to be associated with an AGN-driven galactic outflow \citep{Lamastra2016}.

We conclude that the best way to assess the nature of the acceleration mechanism in the superwind of starbursts is through high-resolution  observations of the non-thermal radiation in the very nearby galaxies NGC 253 and M82 (e.g., with the forthcoming Cherenkov Telescope Array). If the extended gamma-ray emission can be clearly disentangled from the disk contribution and the corresponding spectra well determined, one might favor one of the proposed scenarios over the other.

 \begin{table*}[htbp]
   \caption{Parameters of gamma-ray emitting starbursts.}              
   \label{tab:Lcd}      
   \centering
   \begin{tabular}{l c c c c}          
   \hline\hline                        
   Parameter & \multicolumn{4}{c}{Value}  \\
   & NGC $253$ & M$82$ &  NGC $4945$ & NGC $1068$  \\
   \hline                                   
   Distance $d$ [Mpc] & $2.6$  & $3.4$ & $3.7$ & $16.7$ \\     
   Spectral index $\Gamma$ & $2.2 \pm 0.1$  & $2.2 \pm 0.1$ & $2.1 \pm 0.2$ & $2.2 \pm 0.2$ \\
   $L_{100-300 {\rm MeV}}$ [erg s$^{-1}$] & $1.8 \times 10^{39}$ & $2.5 \times 10^{39}$ & $8.7 \times 10^{39}$ & $8.0 \times 10^{40}$ \\
   $L_{0.3-1 {\rm GeV}}$ [erg s$^{-1}$]  & $1.4 \times 10^{39}$ & $2.4 \times 10^{39}$ & $4.8 \times 10^{39}$ & $4.4 \times 10^{40}$ \\
   $L_{1-3 {\rm GeV}}$ [erg s$^{-1}$]  & $9.8 \times 10^{38}$ & $3.0 \times 10^{39}$ & $2.8 \times 10^{39}$ & $3.6 \times 10^{40}$ \\
   $L_{3-10 {\rm GeV}}$ [erg s$^{-1}$]  & $9.8 \times 10^{38}$ & $ 1.3 \times 10^{39}$ & $2.0 \times 10^{39}$ & $1.9 \times 10^{40}$\\
   $L_{10-100 {\rm GeV}}$ [erg s$^{-1}$] & $4.4 \times 10^{38}$ & $1.2 \times 10^{39}$& $1.1 \times 10^{39}$ & $2.1 \times 10^{40}$\\
   $L_{0.1-100 {\rm GeV}}$ [erg s$^{-1}$]  & $6.0 \pm 0.2 \times 10^{39}$ & $1.5 \pm 0.3 \times 10^{40}$ & $1.2 \pm 0.4 \times 10^{40}$ & $1.5 \pm 0.6 \times 10^{41}$ \\
   \hline                                             
   \end{tabular}
   \end{table*}

\section{Discussion}\label{Discussion}

\subsection{Stretching the limits}

Is there any assumption in the calculations presented so far that might be relaxed to allow for higher energies? The maximum energies obtained for hadrons are limited essentially by the age of the starbursts. We have assumed an age of $\tau=10$ Myr in accordance with the dynamical age of the source. Given the high mass load of the wind, longer durations for the starburst activity seem extremely unlikely \citep{bolatto2013}. At most a factor of a few, which would increase the maximum energies by the same amount in the absence of losses (relevant for SDA), might be added. In such a case, the maximum energies would be limited by diffusive escape, unless special conditions at the shock are assumed.

Another possibility is magnetic field amplification in the presence of a strong shock.  Enhancement of the magnetic field is well established in supernova remnants. Observations of non-thermal X-ray strips and filaments in some remnants allow us to infer values of the magnetic field far above from what is expected in the shocked interstellar medium (e.g., \citealt{bamba2003,berezhko2003,vink2003}). The mechanism for field amplification is not known, but there are several proposals in the literature. Bell instability \citep{bell2004} is perhaps the most popular mechanism, where the amplification is excited by positive currents of cosmic rays propagating upstream from the shock. The instability, however, seems to be mainly efficient in the fast shocks of young supernova remnants. The saturated magnetic field strength is (e.g., \citealt{bustard2017}):
\begin{equation}
  B\approx \sqrt{4\pi P_{\textrm{cr}} \left( \frac{v_{\textrm{shock}}}{c}\right)},
\end{equation}
where $P_{\textrm{cr}}$ is the cosmic ray pressure. The field will be high only in sources with fast shocks and high densities of cosmic rays.  

Turbulent amplification via small-scale dynamo effects is perhaps another possibility. It can operate either in the precursor of the shock \citep{delvalle2016} or in the post-shock region \citep{xu2017}. Amplifications from one to more than three orders of magnitude have been achieved with this mechanism for different setups of initial conditions. 

In the case of NGC 253, in the absence of losses we can express the maximum energy as a function of the magnetic field as:
\begin{equation}
  E_{\textrm{max}}\sim 3.6 \times 10^6 \left( \frac{B}{\mu \textrm{G}}\right)\;\;\; \textrm{GeV}.
\end{equation}
Then, for an amplified field of $B=1$ mG (amplification factor $A=200$), we get:
\begin{eqnarray}
  E^p_{\textrm{max}}&=&3.6\times10^{18} \;\;\textrm{eV} \;\;\;\;\textrm{protons}  \label{Epmax6}\\ 
  E^{\textrm{Fe}}_{\textrm{max}}&=&9.4\times10^{19} \;\;\textrm{eV} \;\;\;\;\textrm{iron nuclei}. \label{Efemax6}
  \end{eqnarray}
  These estimates are much closer to those of \cite{anchordoqui1999}. However, an amplification factor of 200 seems to be energetically impossible for slow shocks. With shock velocities below 1000 km s$^{-1}$ the energy density associated to the ram pressure in a medium of number density 0.01 cm$^{-3}$ is $\sim 10^{-10}$ erg cm$^{-3}$. This is about two orders of magnitude smaller than the necessary magnetic energy density $B^2/8\pi\sim 4\times 10^{-8}$ erg cm$^{-3}$. So either the shock is faster (of the order of $10^{-4}$ km s$^{-1}$) or the medium is two orders of magnitude denser. The former  case seems impossible on the basis of the well-established properties of the wind and its energy budget (see Sect. \ref{Superwinds} and \ref{NGC253}). The latter would require acceleration in local regions of very special conditions, like clumps or other dense inhomogeneities in the superwind. Such a possibility will be investigated elsewhere. Nevertheless, it is worth mentioning here that \textit{if} such extreme amplification of the field occurs somehow,  electrons will cool by synchrotron radiation. The regions of high magnetic field  will appear as  localized non-thermal spots. This will result in X-ray compact sources and X-ray strips such as those observed in Tycho and other supernova remnants. Deep \textit{Chandra} observations might reveal such features in the future, clarifying the picture.
 
\subsection{A starving black hole?}

The nature of the nucleus of NGC 253 is uncertain. The central region of the galaxy is obscured by gas and dust. Also, the effects of stellar winds affect the determination of rotation curves. Historically, the nucleus has been associated with a strong compact radio source dubbed TH2 (after \citealt{turner1985}). \cite{weaver2002} detected hard X-ray emission from this source and suggested that it might be a low-luminosity active galactic nucleus (LLAGN). The source does not have, however, any infrared, optical, or soft X-ray counterpart. A young supernova remnant similar to the Crab located at the distance of the center of NGC 253 should be detectable at IR/optical wavelengths, so it seems to be ruled out as an alternative possibility. The absence of counterparts led \cite{fernandez2009} to propose that TH2 is a starved black hole similar to the Galactic center SgrA$^{*}$. A reanalysis of \textit{Chandra} observations by \cite{muller-sanchez2010} showed the hard X-ray source is actually located at $\sim0''.7$ southwest from TH2 and is not related to the radio source. This source, called X-1, was suggested to be a hidden LLAGN similar to but weaker than the one detected in the starburst galaxy NGC 4945 (see \citealt{muller-sanchez2010} for a full discussion). Recent near IR observations performed with the Gemini South telescope led \cite{gunthardt2015} to suggest a new candidate for the nucleus: the IR peak known as IRC, that is coincident with the radio source TH7 and with a massive star cluster of $1.4\times10^{7}$ $M_{\odot}$. Such a cluster can hide a $\sim10^{6}$ $M_{\odot}$ starved black hole. 

If a black hole rotates in an external poloidal magnetic field, the field lines are dragged by the ergosphere and a potential drop is created in the magnetosphere. If $B$ is the ordered poloidal field near the hole, $h$ the height of the gap, and $a$ the black hole spin, the induced electromotive force is (e.g., \citealt{znajek1978,levinson2000}):
\begin{equation}
  \Delta V\sim 4.5 \times 10^{17} \left(\frac{a}{M}\right)  \left( \frac{h}{R_{\textrm{g}}}\right)^2  \left( \frac{B}{10^4 \textrm{G}}\right)  \left( \frac{M}{10^6\; M_{\odot}}\right)\;\;\; \textrm{V},  \label{V}
 \end{equation}
 where $R_{\textrm{g}}$ is the gravitational radius of the black hole. The energy density of the magnetic field near the horizon is expected to be in equipartition with the energy density of the accreting matter \citep{boldt1999,levinson2002}. This assumption leads to:
 \begin{equation}
    \left( \frac{B}{10^4 \textrm{G}}\right) = 61\; \dot{m}^{1/2}   \left( \frac{M}{10^6\; M_{\odot}}\right)^{-1}, \label{B}
  \end{equation}
where $\dot{m}$ is the accretion rate in Eddington units ($\dot{M}_{\textrm{Edd}}=L_{\textrm{Edd}} c^2$ with $L_{\textrm{Edd}}\approx 1.3\times10^{44} (M/10^{6}\; M_{\odot})$ erg s$^{-1}$). 

If there is a black hole of $10^{6}$ $M_{\odot}$ in NGC 253, its accretion luminosity cannot be larger than the X-ray luminosity inferred from the observations for the central source \citep{muller-sanchez2010}: $L_{\textrm{X}}\sim 10^{40}$ erg s$^{-1}$. Adopting an efficiency of $10$\% (e.g., \citealt{romero-vila2014}), an upper limit $\dot{m}\sim10^{-3}$ is obtained.  Then, according to Eq. (\ref{B}), the magnetic field is $B\sim 2\times 10^4$ G. Taking $a\sim M$ and $h\sim R_{\textrm{g}}$, Eq. (\ref{V}) gives:
\begin{equation}
  \Delta V \sim 9 \times 10^{17} \;\; \textrm{V}.
\end{equation}
This voltage can accelerate charged particles to maximum energies of    
\begin{eqnarray}
  E^p_{\textrm{max}}&=& e\Delta \textrm{V}\approx 9\times10^{17} \;\;\textrm{eV} \;\;\;\;\textrm{protons}  \label{Epmax4}\\ 
  E^{\textrm{Fe}}_{\textrm{max}}&=&  eZ \Delta \textrm{V} \approx 2.3\times10^{19} \;\;\textrm{eV} \;\;\;\;\textrm{iron nuclei}. \label{Efemax4}
  \end{eqnarray}
  
 The actual energies that can be achieved are limited by radiation losses during the acceleration in the gap. Curvature losses, in particular, can be important \citep{levinson2000}, resulting in $\gamma$-ray production. Photopair and photomeson losses might also play some role, especially for more massive black holes (e.g., \citealt{moncada2017}). In the case of a black hole of $10^6$ $M_{\odot}$ these cooling channels are negligible \citep{levinson2002}. 
 
 For lower accretion rates, the maximum energies fall even below what we have obtained in the different cases of diffusive acceleration. For instance, if $\dot{m}\sim 10^{-6}$, $B\sim610$ G and then,
 \begin{eqnarray}
  E^p_{\textrm{max}}&\approx& 2.7\times10^{16} \;\;\textrm{eV} \;\;\;\;\textrm{protons}  \label{Epmax4}\\ 
  E^{\textrm{Fe}}_{\textrm{max}}&\approx& 7.1\times10^{17} \;\;\textrm{eV} \;\;\;\;\textrm{iron nuclei}. \label{Efemax4}
  \end{eqnarray}
 
 The total power extracted from the black hole is:
 \begin{equation}
  L_{\textrm{BH}}\sim 10^{40} \left(\frac{a}{M}\right)^{2} \left( \frac{B}{10^4 \textrm{G}}\right)^{2}  \left( \frac{M}{10^6\; M_{\odot}}\right)^{2}\;\;\; \textrm{erg\;s}^{-1}.  \label{LBH}
 \end{equation} 
 Only a fraction $\alpha_{\textrm{CR}}$ of this power goes into cosmic rays: 
  \begin{equation}
  L_{\textrm{CR}}= \alpha_{\textrm{CR}}\, L_{\textrm{BH}}.
  \end{equation}  
  In terms of the Eddington luminosity, this can be rewritten as: 
  \begin{equation}
  L_{\textrm{CR}}= 4 \times10^{43} \, \alpha_{\textrm{CR}}\, \dot{m} \;\;\;\;  \textrm{erg\;s}^{-1}, \;\;\;\;\;\;\;   \alpha_{\textrm{CR}}<< 1. \label{LCR}
    \end{equation}

If the losses result in a $\gamma$-ray luminosity of $L_{\gamma}=\kappa  L_{\textrm{CR}}$, we can use $\gamma$-ray observations to impose an additional constraint. The integrated flux above 200 MeV of NGC 253 is \citep{abramowski2012} $L_{\gamma}\sim 4.3 \times 10^{39}$ erg s$^{-1}$ (assuming a distance of 2.6 Mpc). Then, Eq. (\ref{LCR}) with $\dot{m}\sim 10^{-3}$ implies $\alpha_{\textrm{CR}}\kappa\sim 0.1$.  For a radiative efficiency $\kappa\sim 1$, we get $\alpha_{\textrm{CR}}\sim 10^{-1}$ and $ L_{\textrm{CR}}\sim 4 \times10^{39}$ erg s$^{-1}$.  This is similar to the luminosity obtained with SDA and two orders of magnitude less than with DSA (see Sect. \ref{sh-lum}).

\subsection{Other potential sources of CR in the disk}

In the starburst region of the galaxy, the enhanced star-forming rate can lead to a number of potential sources of ultra-high energy CRs such as magnetars and mildly relativistic SNe as SNe Ibc, which are more abundant  than the classical long GRBs (e.g., \citealt{Fang2012}, \citealt{Chakraborti2011}). The detection of such sources in the electromagnetic sector, however, is difficult because of the high absorption in the inner galaxy.

The existence of collective wind effects, otherwise, is indubitable.  The interaction of wind-inflated superbubbles with SNe can result is CR acceleration beyond 100 PeV \citep{Bykov2001,Bykov2014,Bykov2018}. These CRs might be convected by the superwind into the halo region of the galaxy, where some reacceleration might occur at the terminal shock. Adiabatic losses, however, should be important for these particles and it is far from clear whether a significant flux with energies above $10^{18}$~eV can be produced and then arrive at the earth (see following section).

\subsection{Propagation and effects upon arrival}

Recent models of the Galactic magnetic field have been developed which are
based on a vast number of measurements of Faraday
rotation~\citep{0004-637X-738-2-192}.
~\cite{Jansson:2012pc,2041-8205-761-1-L11} used polarized synchrotron radiation in addition. Nevertheless, the amount of deflection of extragalactic cosmic
rays in the Galactic magnetic field is rather uncertain~\citep{Unger:2017kfh}.

Even in case of assuming nuclei remaining intact without photo-disintegrating
due to the relative proximity of
NGC 253~\citep{KAMPERT201341,1475-7516-2016-05-038}, deflections caused by the
Galactic magnetic field do not permit the attribution of cosmic rays to a region of
a single source at energies derived for the realistic SDA and DSA scenarios; see Eqs.~(\ref{Epmax2})-(\ref{Efemax2}) and Eqs.~(\ref{Epmax3})-(\ref{Efemax3}), respectively.  Energies at which particles
remain restricted to ballistic
trajectories are only possible when stretching the limits using amplified
magnetic fields at the source are. 

The Galactic magnetic field models mentioned above are used
by ~\cite{Erdmann:2016vle} to determine the minimal rigidity for ballistic deflections to be of the
order of $E/Z=6\;\textrm{EV}$; a factor of two larger than given in
Eq.~(\ref{Epmax6}). In particular, the region at latitudes below $-19.5^\circ$
gives rise to large bulk deflections of up to $50^\circ$ for a rigidity of $E/Z=6\;\textrm{EV}$.  The
identification of NGC 253 as a starburst galaxy emitting cosmic rays at these rigidities seems therefore difficult to achieve, even if such acceleration actually occurs\footnote{A different view based upon \cite{anchordoqui1999} is expressed by \cite{anchordoqui2018}.}.  

Given the increasingly
heavier average mass of ultra-high-energy cosmic rays from proton-dominated
near the ankle region at $10^{18.33}$\;eV to intermediate masses at
$10^{19.65}$\;eV~\citep{Auger2017c,Auger2014}  the search for multiplets of
individual mass groups in the same $E/Z$ range might be
promising~\citep{Abreu:2011md}, but a more detailed assessment of the feasibility
needs to be performed and will be given in a forthcoming paper.

\section{Conclusions}\label{Conclusions}

In this paper we have assessed the potential of starburst galaxies as cosmic ray accelerators. We adopted as a case of study the southern galaxy NGC 253. This class of objects are interesting as cosmic ray sources for several reasons: 1) They are non-thermal radio and gamma-ray emitters, so cosmic rays up to TeV energies are accelerated in them. 2) The star-forming activity releases large amounts of thermal energy that can power very strong winds on galactic scale. 3) These winds create huge bubbles of hot gas above the galactic disks, The size of these regions is large enough as to contain cosmic rays of the highest energies. 4) Strong shocks are produced in the interaction of the superwind with the external medium. Such shocks, as well as the turbulent plasma in the bubbles, can be part of potential mechanisms that accelerate charged particles up to energies far larger than those inferred from the gamma rays. 5) Starburst galaxies are objects of high metallicity, so acceleration of heavy nuclei is favored in them. 

Motivated by these considerations, and after using current multi-wavelength observations to characterize the potential acceleration of particles, we conclude that starburst galaxies in general and NGC 253 in particular might produce cosmic rays with energies up to $10^{18}$ eV in the superwind region, either by diffusive shock acceleration in the reverse terminal shock of the superwind or by stochastic diffusive acceleration in the turbulent gas of the bubbles. We conclude that under normal circumstances acceleration up to 100 EeV is unlikely. The high mass load of the wind results in moderate shock velocities of less than $1000$ km s$^{-1}$ that avoid efficient acceleration on timescales compatible with the age of the starburst episode. The low Alfv\'en velocities in the turbulent plasma of the bubbles, on the other hand, limits the effectiveness of stochastic acceleration. 

Amplified magnetic fields close to the shock, as observed in supernova remnants, might offer a way to reach energies comparable to those of the more energetic cosmic rays. The shocks and the conditions around them, however, are very different in starbursts galaxies from those found in young supernova remnants, so a separate study should be devoted to this issue. In particular, much higher ram pressures (and hence densities) should occur than what we can presently infer from the observations. We notice here that if fields of the order of 1 mG exist in some parts of the gas shocked by the superwind, features in X-rays and radio should exist, and they may be detectable. This offers a way of testing such a hypothesis, which is otherwise based on a poorly established physical basis. 


We conclude that starbursts galaxies are fascinating astrophysical systems, capable of accelerating cosmic rays and producing non-thermal radiation beyond what is seen in our Galaxy. Whether they are actually related to the highest cosmic rays observed on Earth, remains unclear.   

\section*{Acknowledgments}

We thank M. Unger for discussions and an anonymous reviewer for valuable suggestions. GER is very grateful to the IKP at KIT where this research was done. This work was supported by the Helmholtz Association through a Helmholtz International Fellow Award to GER. Additional support was provided by the Argentine agency CONICET (PIP 2014-00338) and the Spanish Ministerio de Econom\'{i}a y Competitividad (MINECO/FEDER, UE) under grant  AYA2016-76012-C3-1-P. 

\section*{Appendix: Adiabaticity and other requirements for efficient particle acceleration at the shocks}
  
  As mentioned in Section \ref{Intro}, the interaction of the superwind with the halo and swept-up disk material produces two shock waves. One of them propagates through the halo/thick disk with velocity $v_{\rm shell}$ whereas the reverse shock moves through the superwind material with a velocity $v_{\rm rev}$. In our model, for thermalization $\epsilon=0.75$, these velocities are  \mbox{$298$ km s$^{-1}$} and      
  \mbox{$750$ km s$^{-1}$}, respectively, and $328$ km s$^{-1}$ and $866$ km s$^{-1}$ for $\epsilon \approx 1$. Using these values and the number density of each pre-shocked medium, $n_{\rm sw}=2 \times 10^{-3}$ cm$^{-3}$ for the superwind bubble and $n_{\rm halo/disk}=6.8 \times 10^{-3}$ cm$^{-3}$ for the halo/swept-up material \citep{strickland2002}, we study the nature of these shocks. For this purpose, we calculate the cooling length $R_{\Lambda}$ using the following expression \citep{mccray1979}:

   \begin{equation}
    R_{\Lambda}=\frac{1.90\times10^{-29}(v_{\textrm{s}}/\textrm{km s$^{-1}$})^{3}\, \mu}{(n/ \textrm{cm$^{-3}$})\,(\Lambda(T)/\textrm{erg cm$^{3}$ s$^{-1}$})}\textrm{ pc}
   \label{eqn:thermallength}
   \end{equation}
   
   \begin{equation}
    \textrm{ with } \,\,\, T=18.21\, \mu \, \left( \frac{v_{\textrm{s}}}{\textrm{km s$^{-1}$}} \right)^{2} \text{ K,}
   \end{equation}

  \noindent where $\mu$ is 0.6 if the material is ionized or 1.3 if neutral, $v_{\textrm{s}}$ is the shock velocity, $n$ is the number density of the non-pertubated medium, and $\Lambda(T)$ [erg cm$^{3}$ s$^{-1}$] is the cooling function \citep{raymond1976,myasnikov1998}:
  
  \begin{equation}
   \Lambda(T)=\left\{
               \begin{array}{lll}
                 7\times 10^{-27}T  & \mathrm{if\ } 10^{4}\,{\rm K} \le T \le 10^{5}\,{\rm K} \\
                 7\times 10^{-19}T^{-0.6} & \mathrm{if\ } 10^{5}\,{\rm K} \le T \le 4\times 10^{7}\,{\rm K} \\
                 3\times 10^{-27}T^{0.5} & \mathrm{if\ } T \ge 4\times 10^{7}\,{\rm K}  \\
               \end{array}
             \right.
  \end{equation}
  
  The adiabaticity of the shock can be determined comparing this cooling length with the length traversed by the shock. If the cooling length is longer, the shock is adiabatic, otherwise it is radiative. An equivalent analysis is possible using the timescale of the thermal cooling  instead of the cooling length:
  
  \begin{equation}
   t_{\rm th} \approx \frac{4\,R_{\Lambda}}{v_{\textrm{s}}}
  ,\end{equation}
 
  \noindent and comparing it with the lifetime of the source. Table \ref{tab:adiabaticity} shows the results for the values of the parameter  in our model. 
  
  \begin{table*}[ht]
    \caption{Parameters for the forward and reverse shocks.}
        \label{tab:adiabaticity}
        \centering
   \begin{tabular}{lll}
    \hline\hline %
     Parameters & \multicolumn{2}{c}{$\epsilon=1$}\\
     & Forward shock & Reverse shock\\
    \hline
    $v$ [km s$^{-1}$] & $328$ & $866$\\
    $\mu$ & $0.60$ & $0.60$ \\
    $T$ [K] & $1.2 \times 10^{6}$ & $8.20 \times 10^{6}$ \\
    $n$ [cm$^{-3}$]  &$6.83 \times 10^{-3}$ & $2 \times 10^{-3}$\\
    $t_{\textrm{th}}$ [yr]  &$3 \times 10^{6}$ & $2.24 \times 10^{8}$ \\
       \hline\hline %
     Parameters & \multicolumn{2}{c}{$\epsilon=0.75$}\\
     & Forward shock & Reverse shock\\
    \hline
    $v$ [km s$^{-1}$] & $298$ & $750$\\
    $\mu$ & $0.60$ & $0.60$ \\
    $T$ [K] & $10^{6}$ & $6.10 \times 10^{6}$ \\
    $n$ [cm$^{-3}$] &$6.83 \times 10^{-3}$ & $2 \times 10^{-3}$\\
    $t_{\textrm{th}}$ [yr]  &$2.16 \times 10^{6}$ & $1.40 \times 10^{8}$\\
    \hline \\
  \end{tabular}                                                                                                                                                                         
  \end{table*} 
  
  Since the lifetime of the source is $\sim 10^{7}$ yr in our model, the forward shocks turn out to be radiative and the reverse shocks adiabatic for both thermalization efficiencies, as we mentioned in the main body of this work.
  
 Another condition that must be satisfied by a shock in order to accelerate particles is that the overall shock Mach number exceeds the critical value $M>\sqrt{5}$ \citep{vink2014}. As we can see from Table \ref{T1}, the Mach numbers are $M\approx 5.3$ and $M\approx 4.6$, for the thermalization parameters $\epsilon=1$ and   $\epsilon=0.75$, respectively. The condition, therefore, is largely fulfilled. 

\bibliographystyle{aa}
\bibliography{myrefs8}   

\label{lastpage}

\end{document}